\documentclass[trackchanges,twocolumn]{aastex7}
\DeclareSymbolFont{matha}{OML}{txmi}{m}{it}
\DeclareMathSymbol{\varv}{\mathord}{matha}{118}

\newcommand{\eps}{\epsilon}
\newcommand{\epscrit}{\eps_{\rm{crit}}}

\newcommand{\Zcrit}{Z_{\rm{crit}}}

\newcommand{\rhopmax}{\rho_{p,\rm{max}}}
\newcommand{\rhoH}{\rho_{\rm{H}}}

\usepackage{natbib}
\usepackage{amsmath}
\usepackage{amssymb}
\usepackage{bm}
\usepackage{comment}
\usepackage[T1]{fontenc}

\defcitealias{carrera_how_2015}{C15}
\defcitealias{Yang2017}{Y17}
\defcitealias{LiYoudin21}{LY21}
\defcitealias{Lim25}{L25}
\usepackage{CJKutf8}

\usepackage{hyperref}

\shorttitle{SI in 3D}
\shortauthors{Lim et al.}

\begin{document}
\begin{CJK}{UTF8}{mj}
\title{The Streaming Instability in 3D: Conditions for Strong Clumping}

\author[orcid=0000-0003-2719-6640]{Jeonghoon Lim (임정훈)}
\affiliation{Department of Physics and Astronomy, Iowa State University, Ames, IA 50010, USA}
\email{jhlim@iastate.edu}

\author[orcid=0000-0002-3771-8054]{Jacob B. Simon}
\affiliation{Department of Physics and Astronomy, Iowa State University, Ames, IA 50010, USA}
\email{jbsimon@iastate.edu}

\author[orcid=0000-0001-9222-4367]{Rixin Li (李日新)}
\altaffiliation{51 Pegasi b Fellow}
\affiliation{Department of Astronomy, University of California, Berkeley, Berkeley, CA 94720, USA}
\email{rixin@berkeley.edu}

\author[orcid=0009-0006-2478-5246]{Olivia Brouillette}
\affiliation{Department of Astronomy, New Mexico State University, PO Box 30001 MSC 4500, Las Cruces, NM 88001, USA}
\email{ob1@nmsu.edu}

\author[orcid=0000-0002-5000-2747]{David G. Rea}
\affiliation{Department of Physics and Astronomy, Iowa State University, Ames, IA 50010, USA}
\email{drea1@iastate.edu}

\author[orcid=0000-0002-3768-7542]{Wladimir Lyra}
\affiliation{Department of Astronomy, New Mexico State University, PO Box 30001 MSC 4500, Las Cruces, NM 88001, USA}
\email{wlyra@nmsu.edu}

\correspondingauthor{Jeonghoon Lim}
\email{jhlim@iastate.edu}

\begin{abstract}
The streaming instability (SI) is a leading mechanism for planetesimal formation, driving the aerodynamic concentration of solids in protoplanetary disks. The SI triggers strong clumping (i.e., strong enough for clumps to collapse) when the solid-to-gas column density ratio, $Z$, exceeds a threshold, $\Zcrit$. This threshold depends on the dimensionless stopping time, $\tau_s$. Although the strong-clumping threshold has been explored over the last decade, it has been determined largely through 2D axisymmetric simulations. In this work, we perform a suite of 3D, vertically stratified simulations to establish a clumping threshold across $10^{-3} \leq \tau_s \leq 1.0$. Additionally, we study SI-driven concentration that is unique to 3D. We find that $\Zcrit$ is as low as $\approx 0.002$ at $\tau_s=0.1$ and exceeds $\approx 0.03$ at $\tau_s=10^{-3}$. Compared to 2D, our 3D results yield lower $\Zcrit$ for $\tau_s > 0.02$, but higher for $\tau_s \leq 0.02$, with a sharp transition between $\tau_s = 0.02$ and 0.03. This transition correlates with midplane density ratio ($\epsilon$): $\epsilon < 1$ where 3D gives lower thresholds, and $\epsilon > 1$ where 3D gives higher thresholds. We also find a filaments-in-filaments structure when $\epsilon < 1$, which enhances clumping compared to 2D. By contrast, when $\epsilon > 1$ and $\tau_s \leq 0.03$, dust filaments in 3D do not drift inward, suppressing filament mergers and strong clumping. In 2D, filaments drift inward regardless of $\epsilon$, triggering strong clumping easier in this regime. Our results underscore the necessity of 3D simulations for accurately capturing SI-driven concentration and building the strong-clumping threshold.
\end{abstract}


\section{Introduction}\label{sec:intro}
Growth from micron-sized dust grains to kilometer-sized planetesimals marks a crucial first step in planet formation (see \citealt{Simon2024review} for the review). Dust grains initially grow through collisional coagulation, but as they increase in size and decouple from the gas, their relative velocities rise, leading to destructive collisions. In addition, larger grains experience faster radial drift toward the star \citep{Weidenschilling77}. Although the exact outcome depends on disk conditions (e.g., turbulence) and grain properties (e.g., composition and porosity), previous studies have shown that both fragmentation and radial drift barriers significantly suppress grain growth, with the fragmentation barrier limiting sizes to centimeters \citep{PPVII_planet_formation}. Overcoming these barriers requires alternative mechanisms to bridge the gap from centimeter-sized solids to planetesimals.

One way to bypass the growth barriers and to form planetesimals is through the direct gravitational collapse of dense clouds of solids. For collapse to proceed, the local concentration of solids must be strong enough to overcome tidal shear and turbulent diffusion \citep{Schreiber_Klahr2018,Gerbig20,gerbig_planetesimal_2023}. One pathway for achieving such concentrations is dust trapping at pressure maxima \citep{Johansen06,Carrera2022,schafer_coexistence_2022,xu_dust_2022} or within disk vortices \citep{Lyra24}, both of which can significantly enhance the local solid density. Another mechanism is the streaming instability (SI; \citealt{YG05,YJ07,JY07}), a linear instability that arises in rotating dust-gas mixtures coupled through aerodynamic drag in the presence of a radial pressure gradient—conditions that naturally present in PPDs.

The SI has attracted significant attention in numerical studies due to its ability to form dense clumps of solids that can gravitationally collapse into planetesimals. For example, previous numerical simulations have demonstrated that under certain conditions, the SI efficiently concentrates solids into narrow filaments on scales smaller than the gas pressure scale height, thereby greatly enhancing the local dust-to-gas density ratio (e.g., \citealt{JY07, BaiStone10b_stratified}). Numerical simulations that incorporate self-gravity of solids have shown that planetesimals form within such dense filaments \citep{johansen_rapid_2007, johansen15, simon_mass_2016, li_demographics_2019, abod_mass_2019,Schafer2024}. Analyses of the resulting planetesimals indicate that their properties are consistent with some observational constraints: the inclination and angular momentum distributions of Kuiper Belt objects \citep{Nesvorny2019, Nesvorny2021} and the observed correlation between planetesimal densities and sizes \citep{Canas24}.

The propensity of the SI to clump solids is largely influenced by two key parameters. The first is the dimensionless stopping time, $\tau_s$, which represents the timescale—measured relative to the local Keplerian frequency $\Omega$—over which the relative velocity between solids and gas decays due to gas drag. This parameter is proportional to the size of solid particles. The second parameter is $Z/\Pi$ \citep{sekiya_two_2018}, where $Z$ is the solid-to-gas column density ratio, and $\Pi$ measures the deviation of rotational velocity from Keplerian in units of local sound speed due to a radial pressure gradient (see Section~\ref{sec:methods}). This ratio estimates the column density ratio within the dust layer instead of across the full vertical extent of the disk, since the dust layer thickness scales with $\Pi$ \citep{Baronett24,Lim25_unstrat_strat}. For example, numerical simulations have shown that the same $Z/\Pi$ but different individual values of $Z$ and $\Pi$ yield similar behaviors of the SI in its saturated state. \citep{sekiya_two_2018,Lim25_unstrat_strat}. Though, it is worth noting that how the saturated state of the SI (as set by $Z/\Pi$)  determines strong clumping (see Section~\ref{sec:methods} for its definition) is unclear \citep{Lim25_unstrat_strat}.

Several previous studies have investigated the conditions under which SI-driven concentration becomes strong enough to trigger planetesimal formation, typically characterized by combinations of $\tau_s$ and $Z$ with $\Pi$ fixed to 0.05 (but see \citealt{LiYoudin21}, hereafter \citetalias{LiYoudin21}, for how the clumping condition can be generalized for nonzero $\Pi$ values). This effort was initiated by \citet[][hereafter \citetalias{carrera_how_2015}]{carrera_how_2015}, who identified a U-shaped clumping boundary in the $\tau_s$–$Z$ parameter space. Subsequent works have improved the clumping boundary by using improved numerical setups, such as an improved algorithm for stiff drag force (\citealt{Yang_Johansen16,Yang2017}, hereafter \citetalias{Yang2017}), larger domain sizes (\citetalias{Yang2017}; \citealt{LiYoudin21}, hereafter \citetalias{LiYoudin21}), and grid resolution (\citetalias{Yang2017,LiYoudin21}; \citealt{Lim25}, hereafter \citetalias{Lim25}). 

The most up-to-date version of the clumping boundary was established by \citetalias{Lim25}. They employed higher grid resolutions than any previous study and performed simulations over longer durations—both critical for small $\tau_s$, as smaller values lead to shorter wavelengths of the fastest-growing SI modes \citep{YJ07} and require longer timescales for filament formation and merging \citepalias{Yang2017}. As a result, they lowered the clumping threshold at $\tau_s = 10^{-2}$ from above $Z = 0.01$ (as reported by \citetalias{Yang2017} and \citetalias{LiYoudin21}) to below 0.01. Their results demonstrated that efficient planetesimal formation via the SI is possible under less stringent conditions than previously thought. 

However, all the aforementioned studies on the clumping boundary were largely based on 2D axisymmetric (radial–vertical) simulations.\footnote{We note that \citetalias{Yang2017} also performed 3D simulations in addition to 2D, but their 3D runs were limited to only two $(\tau_s, Z)$ combinations, compared to six explored in 2D. In addition, although the clumping threshold in \citetalias{LiYoudin21} was derived from 2D models, they performed a single 3D run that indicated that 3D simulations may lower the clumping threshold.} By contrast, dust–gas dynamics in 3D are inherently non-axisymmetric. Several non-axisymmetric effects—such as small-scale clumping \citepalias{Yang2017}, outward migration of filaments \citepalias{Yang2017}, the development of a filaments-in-filaments structure \citepalias{LiYoudin21}—have been previously reported and are also observed in our 3D simulations. Moreover, \citet{Schafer2024} showed that the maximum azimuthal extent of filaments is approximately one gas scale height, indicating that the SI produces a network of loosely connected filaments rather than axisymmetric, ring-like structures.

Motivated by the significant influence of the azimuthal dimension, we aim to further refine the clumping boundary by conducting 3D simulations using the shearing-box approximation with vertical gravity. We explore a broad range of stopping times ($\tau_s = 10^{-3}$ to 1.0) and find that the resulting clumping boundary differs substantially from that derived in previous 2D studies.

The paper is organized as follows: Section~\ref{sec:methods} describes our numerical setup and simulation parameters. Section~\ref{sec:results_3D} presents the main results from our 3D simulations, including the updated clumping boundary (see Section~\ref{sec:results_3D:crit_ratio}).  In Section~\ref{sec:results:comp_2D}, we perform systematic comparisons between 2D and 3D simulations for selected $(\tau_s, Z)$ combinations to illustrate the effect of non-axisymmetry. We provide discussion and summary in Sections~\ref{sec:discussion}-\ref{sec:summary}, respectively. 

\begin{deluxetable*}{ccccccccccc}
\tablecaption{List of Simulations and Time-averaged Quantities}
\label{table:simulations}
\tablehead{
\colhead{$\tau_s$} & 
\colhead{$Z$} & 
\colhead{$L_x\times L_y\times L_z$} & 
\colhead{Resolution} & 
\colhead{$\rho_{p,\rm{max}}$} & 
\colhead{$\epsilon$} & 
\colhead{$\langle H_p^* \rangle$} & 
\colhead{$H_p$} & 
\colhead{$[t_s, t_e]$} & 
\colhead{$t_{\rm{end}}$} &
\colhead{Strong clumping} \\ 
\colhead{} &
\colhead{} &
\colhead{$H^3$} &
\colhead{$H^{-1}$} &
\colhead{$\rho_{g0}$}  &
\colhead{}  &
\colhead{$H$} &
\colhead{$H$} &
\colhead{$\Omega^{-1}$} &
\colhead{$\Omega^{-1}$} &
\colhead{}  \\
\colhead{(1)} &
\colhead{(2)} &
\colhead{(3)} &
\colhead{(4)} &
\colhead{(5)} &
\colhead{(6)} &
\colhead{(7)} &
\colhead{(8)} &
\colhead{(9)} &
\colhead{(10)} &
\colhead{(11)} 
}

\startdata
0.001 & 0.03 & $0.4\times 0.2\times0.2$ & 640 & 12.289 & 5.276 & 0.006 & 0.006 & [1200,3000] & 3000 & N \\ 
0.01 & 0.02 & $0.4\times 0.2\times0.2$ & 640 & 7.210 & 3.213 & 0.006 & 0.006 & [200,3000] & 3000 & N \\ 
0.01 & 0.03 & $0.4\times 0.2\times0.2$ & 640 & 23.404 & 5.380 & 0.006 & 0.006 & [200,2394] & 2486\tablenotemark{a} & Y \\
0.02 & 0.006 & $0.8\times 0.2\times0.2$ & 640 & 1.895 & 0.971 & 0.006 & 0.006 & [150,3000] & 3000 & N \\
0.02 & 0.0075 & $0.8\times 0.2\times0.2$ & 640 & 2.687 & 1.179 & 0.006 & 0.006 & [150,3000] & 3000 & N \\
0.02 & 0.01 & $0.8\times 0.2\times0.2$ & 640 & 3.663 & 1.545 & 0.007 & 0.007 & [150,3000] & 3000 & N \\
0.02 & 0.013 & $0.8\times 0.2\times0.2$ & 320 & 7.651 & 3.021 & 0.004 & 0.004 & [150,3000] & 3000 & N \\
0.02 & 0.013 & $0.8\times 0.2\times0.2$ & 640 & 7.506 & 2.030 & 0.007 & 0.007 & [150,5000] & 5000 & N \\
0.02 & 0.013 & $0.8\times 0.2\times0.2$ & 1280 & 4.684 & 1.619 & 0.008 & 0.008 & [150,5000] & 5000 & N \\
0.02 & 0.013 & $0.2\times 0.2\times0.2$ & 2560 & 3.106 & 1.066 & 0.012 & 0.012 & [300,1500] & 1500\tablenotemark{b} & N \\
0.02 & 0.02 & $0.8\times 0.2\times0.2$ & 640 & 24.676 & 2.794 & 0.007 & 0.007 & [150,2997] & 3100 & Y \\
0.03 & 0.002 & $0.8\times 0.2\times 0.2$ & 640 & 0.572 & 0.230 & 0.009 & 0.010 & [100,3000] & 3000 & N \\
0.03 & 0.003 & $0.8\times 0.2\times 0.2$ & 640 & 7.015 & 0.391 & 0.008 & 0.007 & [100,3000] & 3000 & N \\
0.03 & 0.004 & $0.8\times 0.2\times 0.2$ & 640 & 12.712 & 0.577 & 0.007 & 0.007 & [100,3000] & 3000 & N \\
0.03 & 0.005 & $0.8\times 0.2\times 0.2$ & 640 & 9.289 & 0.697 & 0.007 & 0.007 & [100,1737] & 3000 & Y \\
0.03 & 0.006 & $0.8\times 0.2\times 0.2$ & 640 & 7.739 & 0.852 & 0.007 & 0.007 & [100,2323] & 3000 & Y \\
0.03 & 0.02 & $0.8\times 0.2\times 0.2$ & 640 & 14.201 & 2.245 & 0.009 & 0.009 & [100,2560] & 2700\tablenotemark{c} & Y \\
0.05 & 0.0015 & $0.8\times 0.2\times 0.2$ & 640 & 9.548 & 0.195 & 0.008 & 0.008 & [100,3000] & 3000 & N \\
0.05 & 0.002 & $0.8\times 0.2\times 0.2$ & 640 & 14.603 & 0.269 & 0.007 & 0.007 & [100,3000] & 3000 & N \\
0.05 & 0.003 & $0.8\times 0.2\times 0.2$ & 640 & 15.225 & 0.367 & 0.008 & 0.008 & [100,1465] & 3000 & Y \\
0.05 & 0.004 & $0.8\times 0.2\times 0.2$ & 640 & 11.941 & 0.470 & 0.009 & 0.009 & [100,1571] & 3000 & Y \\
0.05 & 0.005 & $0.8\times 0.2\times 0.2$ & 640 & 10.321 & 0.530 & 0.010 & 0.010 & [100,856] & 3000 & Y \\
0.1 & 0.0015 & $0.8\times 0.2\times 0.2$ & 640 & 15.749 & 0.223 & 0.007 & 0.008 & [50,2000] & 2000 & N \\
0.1 & 0.002 & $0.8\times 0.2\times 0.2$ & 640 & 12.935 & 0.251 & 0.008 & 0.009 & [50,270] & 2000 & Y \\
0.1 & 0.003 & $0.8\times 0.2\times 0.2$ & 640 & 8.976 & 0.331 & 0.009 & 0.010 & [50,332] & 2000 & Y \\
0.1 & 0.004 & $0.8\times 0.2\times 0.2$ & 640 & 8.902 & 0.444 & 0.009 & 0.010 & [50,368] & 2000 & Y \\
0.1 & 0.02 & $0.8\times 0.2\times 0.2$ & 640 & 52.696 & 2.632 & 0.008 & 0.008 & [50,213] & 915\tablenotemark{c} & Y \\
0.3 & 0.002 & $0.8\times 0.2\times 0.2$ & 640 & 6.019 & 0.345 & 0.006 & 0.009 & [100,1000] & 1000 & N \\
0.3 & 0.003 & $0.8\times 0.2\times 0.2$ & 640 & 16.461 & 0.475 & 0.006 & 0.009  & [100,210] & 1000 & Y \\
0.3 & 0.004 & $0.8\times 0.2\times 0.2$ & 640 & 15.805 & 0.549 & 0.007 & 0.010 & [100,458] & 1000 & Y \\
1.0 & 0.002 & $0.8\times 0.2\times 0.2$ & 640 & 11.781 & 1.070 & 0.002 & 0.005 & [100,500] & 500 & N \\
1.0 & 0.003 & $0.8\times 0.2\times 0.2$ & 640 & 22.046 & 1.322 & 0.002 & 0.005 & [100,329] & 500 & Y \\
1.0 & 0.004 & $0.8\times 0.2\times 0.2$ & 640 & 20.097 & 1.340 & 0.003 & 0.006 & [100,370] & 500 & Y \\
\enddata
\tablecomments{Columns: (1): dimensionless stopping time of solids (Equation \ref{eq:taus}); (2): surface density ratio of solids to gas (Equation \ref{eq:Z}); (3): domain size in units of $H$; (4): the number of grid cells per $H$; (5)-(8): time-averaged maximum solid density, midplane solid-to-gas density ratio (Equation~\ref{eq:eps}), effective scale height of solids (Equation~\ref{eq:Hpeff}), and the scale height computed from all particles (Equation~\ref{eq:Hp}), respectively (9): start ($t_s$) and end ($t_e$) time of an interval over which the time-averages are done; (10): total simulation time; (11): whether or not the strong clumping occurs (Y for Yes and N for No). All runs have the same global radial pressure gradient set by $\Pi = 0.05$. 
}
\tablenotetext{a}{We terminated this run after strong clumping begins because the stiffness-mitigating technique (see Section \ref{sec:methods}) causes a prohibitive computational cost. We nevertheless count this run as strong clumping because $\rhopmax$ exceeds $\rhoH$ at $t\Omega\approx 2439$ after which it maintains an upward trend until $t_{\rm{end}}$.}
\tablenotetext{b}{The high resolution of this simulation makes it computationally infeasible to evolve it for the same duration as other simulations with the same $\tau_s$. Moreover, $\rhopmax$ saturates at approximately $3\rho_{g0}$ with no indication of an upward trend.}
\tablenotetext{c}{These runs are not intended to determine the strong-clumping threshold--as it occurs at much lower $Z$--but are used for comparison to smaller $Z$ (Section \ref{sec:results_3D:SI_clumping}). Therefore, we evolve these runs only slightly beyond the time at which strong clumping occurs.}
\end{deluxetable*}

\section{Methods} \label{sec:methods}
We perform numerical simulations of the coupled dynamics of the gas and solids in a protoplanetary disk. The numerical method is similar to those of \citetalias{Lim25} (see their Equations~1-4 for the governing equations we use in this work), summarized as follows. 

We use the {\sc Athena} code \citep{Stone08} to model isothermal, unmagnetized gas, coupled with the Lagrangian particle module developed by \citet{BaiStone10a}. Simulations are performed using the local shearing box approximation, which models a vertically stratified, co-rotating patch of a protoplanetary disk \citep{Hawley95, stone_implementation_2010}. The shearing box is centered at a fiducial disk radius $r$, rotates with the corresponding Keplerian frequency $\Omega$, and adopts local Cartesian coordinates ($x, y, z$) representing the radial, azimuthal, and vertical directions, respectively. Our models are fully three-dimensional, in contrast to the 2D axisymmetric ($x$-$z$) simulations presented in \citetalias{Lim25}. We apply standard shear-periodic boundary conditions in the $x$ direction, purely periodic boundaries in the $y$ direction, and outflow boundary conditions in $z$ \citep{Li18}.

Our main goal is to identify the conditions under which the SI leads to strong clumping in a laminar disk—that is, in the absence of external turbulence. Although we do not include self-gravity of solid particles in our simulations, we are interested in whether the SI can produce clumping strong enough to trigger gravitational collapse and planetesimal formation. Thus, following \citetalias{LiYoudin21} and  \citetalias{Lim25}, we classify a simulation as a strong-clumping case when maximum density of solid particles $(\rhopmax)$ exceeds the Hill density, defined as
\begin{equation}\label{eq:rhoH}
    \rhoH=9\sqrt{\frac{\pi}{8}}Q\rho_{g0} \simeq 180\rho_{g0},
\end{equation}
where $Q$ is the Toomre parameter \citep{Safronov1960,Toomre1964}, $\rho_{g0}$ is an initial midplane gas density. We adopt $Q = 32$, representing low-mass disks, to align with previous studies that used the same value and to maintain a conservative threshold for strong clumping. That said, in more massive disks with lower $Q$, strong clumping could occur at weaker concentrations. 

We explore combinations of $(\tau_s,Z)$ to identify conditions for strong clumping, with both parameters described as follows. The dimensionless stopping time of grains $(\tau_s)$ measures how strong they are aerodynamically coupled to the gas, defined as 
\begin{equation}\label{eq:taus}
    \tau_s=t_s\Omega,
\end{equation}
where $t_s$ is the timescale over which particles lose their momentum due to gas drag. We consider only one particle size per simulation, defined via a single value of $\tau_s$ in each simulation, with $\tau_s=10^{-3}$ to 1.0.

The coupled equations for particles and the gas become stiff when $\tau_s$ is small and/or particle-to-gas density ratio is high \citep{BaiStone10a,Yang_Johansen16}. To address this issue, we adopt the stiffness-mitigation technique developed by \citetalias{LiYoudin21} for all runs with $\tau_s \leq 0.01$. In {\sc Athena}, stiffness is quantified by the dimensionless parameter
\begin{equation}\label{eq:stiffness}
    \chi \equiv \frac{(\rho_p/\rho_g) \Delta t_{\rm{CFL}}}{\textrm{max}(\tau_s,\Delta t_{\rm{CFL}})},
\end{equation}
where $\Delta t_{\rm{CFL}} \sim C_0\Delta x/c_s$ and $C_0 = 0.4$ is the Courant-Friedrichs-Lewy (CFL) number in {\sc Athena}. When $\chi \gtrsim 3-5$ (depending on the actual problem),  numerical instabilities can arise, resulting in unphysically large growth of gas and particle velocities and suppressing clumping (\citealt{BaiStone10a}; \citetalias{LiYoudin21}). To prevent this, we reduce the timestep as needed to keep $\chi \lesssim 0.75$ throughout the entire duration of simulations with $\tau_s \leq 0.01$. In addition, we switch to the fully implicit integrators for particles \citep{BaiStone10a} in the run with $\tau_s=10^{-3}$; the semi-implicit integrator \citep{BaiStone10a}  is used in the rest of our simulations. 

The surface density (i.e., vertically integrated density) ratio of solid-to-gas $(Z)$ parameterizes the abundance of solids relative to the gas, defined as
\begin{equation}\label{eq:Z}
    Z=\frac{\Sigma_{p0}}{\Sigma_{g0}}.
\end{equation}
Here, $\Sigma_{p0}$ $(\Sigma_{g0})$ is the initial surface density of solids (the gas). We consider $Z=0.0015$ to 0.03 across simulations. 

In our simulations, a global radial pressure gradient is parameterized by 
\begin{equation}\label{eq:Pi}
   \Pi=\frac{\eta u_K}{c_s}=\frac{\eta r}{H}. 
\end{equation}
Here, $u_K$ is the Keplerian velocity, $c_s$ is sound speed, and $H$ is the vertical scale height of the gas. The parameter $\eta$ is proportional to the radial pressure gradient, quantifying the deviation of the gas rotational velocity from $u_K$ due to pressure gradient support. In our setup, we apply an inward acceleration to particles \citep{BaiStone10a}, which is mathematically equivalent to applying an outward acceleration to the gas. As a result, the gas remains on Keplerian orbits, while particles are azimuthally boosted by $\Pi c_s$ in the absence of backreaction. Furthermore, the parameter $\Pi$ plays a key role in gas-dust dynamics, as it sets the characteristic length and velocity scales of the SI in both linear theory \citep{YG05} and numerical simulations \citep{sekiya_two_2018,Baronett24,Lim25_unstrat_strat}. Nevertheless, since our primary focus is on identifying strong clumping conditions in terms of $\tau_s$ and $Z$, and given that $\Pi = 0.05$ is commonly adopted in previous clumping threshold studies \citepalias{carrera_how_2015,Yang2017, LiYoudin21, Lim25}, we fix $\Pi$ at 0.05 across all simulations. 

Our fiducial grid resolution is 640 cells per $H$, corresponding to a grid cell size of $\Delta x = H/640 \approx 1.6 \times 10^{-3}H$. To assess the effect of resolution on the clumping threshold, we perform higher-resolution tests—doubling and quadrupling the resolution to 1280 and 2560 cells per $H$, respectively—for the $\tau_s = 0.02,~ Z = 0.013$ case (see Section \ref{sec:results:comp_2D:resolution}). 

Most simulations use domain sizes of $(L_x, L_y, L_z) = (0.8, 0.2, 0.2),H$, where $L_x$, $L_y$, and $L_z$ are the radial, azimuthal, and vertical dimensions of the computational domain, respectively. For runs with $\tau_s \leq 0.01$, we reduce the radial domain size to $0.4H$ to offset the high computational cost introduced by the stiffness-mitigation technique. A radially wide domain is used to promote formation of multiple SI filaments and their interactions along the radial directions, as demonstrated in previous studies (e.g., \citealt{YangJohansen14,Li18}). Nevertheless, $L_x=0.4H$ has been shown to accommodate multiple filaments for $\tau_s \leq 0.01$ \citepalias{Yang2017,LiYoudin21,Lim25}.

We assign, on average, one particle per grid cell in all simulations. Thus, the total number of particles $(N_{\rm{par}})$ is  
\begin{equation}\label{eq:Npar}
    N_{\rm{par}}=n_pN_xN_yN_z \approx 8.4\times 10^6\frac{L_xL_yL_z}{(0.8\times0.2^2)H^3},
\end{equation}
where $n_p$ is the average number of particles per cell, and $N_{x,y,z}$ are the number of grid cells in the radial, azimuthal, and vertical directions. The fiducial particle resolution $(n_p=1)$ was shown to be enough to properly capture the density distribution of particles in vertically unstratified simulations \citep{BaiStone10a}. In our vertically stratified setup, $n_p > 1$ near the midplane, with the actual value depending on the thickness of the dust layer, typically on the order of $\Pi H$. Therefore, our particle resolution is expected to be adequate for resolving the particle density distribution in the midplane region.

We initialize the gas with a Gaussian density profile with scale height $H$. Similarly, particles are initially distributed with a Gaussian density profile with an initial scale height of $0.025H$ in the vertical direction. As the simulation evolves, particles settle toward the midplane but are subsequently stirred by turbulence, eventually reaching a quasi-steady state in which vertical settling is balanced by turbulent diffusion. Since we have verified that $\rho_p(z)$ maintains an approximately  Gaussian profile in all of our simulations, the scale height of particles ($H_p$) can be approximated by the standard deviation of particles' vertical position $(z_p)$: 
\begin{equation}\label{eq:Hp}
    H_p \simeq \sqrt{\overline{z_p^2}-(\overline{z_p})^2},
\end{equation}
where the overbar denotes an average over all particles. However, Equation~\eqref{eq:Hp} overestimates a dust scale height when the dust layer is strongly undulated. The undulation becomes more pronounced as $\tau_s \rightarrow 1$ in our simulations (see Section \ref{sec:results_3D:dust_layer}; see also \citetalias{LiYoudin21}). To remove this effect from the scale height measurement, we measure the effective particle scale height ($H_p^*$) by 
\begin{equation}\label{eq:Hpeff}
    H_p^*(x,y) \equiv \sqrt{\frac{\int \rho_p z^2dz}{\int \rho_pdz} - \left(\frac{\int \rho_p zdz}{\int \rho_pdz}\right)^2}.
\end{equation}
Throughout this paper, we use $H_p^*$ for the measurement of particle scale height, and we report temporal averages of $H_p$ and $\langle H_p^* \rangle$ (where $\langle \cdots \rangle$ denotes a spatial average) in Table \ref{table:simulations}. 
Using $H_p^*$, we estimate the midplane solid-to-gas density ratio\footnote{Since $H_p^*$ measures the vertical dispersion of particles about their mean position at each radial position, the density ratio obtained by Equation~\eqref{eq:eps} may reflect off-midplane values, especially when the dust layer undergoes significant vertical undulation, as seen in our simulations with $\tau_s = 0.3$ and 1.0. However, because the dust layer remains extremely thin at these $\tau_s$ values, the computed $\epsilon$ still closely approximates the midplane dust-to-gas density ratio.} by
\begin{equation}\label{eq:eps}
    \epsilon \equiv \frac{Z}{\langle H_p^* \rangle/H}. 
\end{equation}

The vertical velocities of the gas and particles are initially zero. In the radial and azimuthal directions, the positions of particles are randomly chosen from a uniform distribution. We apply the Nakagawa—Sekiya—Hayashi (NSH) equilibrium solutions \citep{Nakagawa1986} to the horizontal velocities (i.e., radial and azimuthal) of the gas and particles initially.  

List of simulations is shown in Table \ref{table:simulations}. We evolve our 3D models for longer duration for smaller $\tau_s$ (see $t_{\rm{end}}$ for each run in the table). This is because the SI is expected to develop after sedimentation has completed and an equilibrium dust layer has formed—a process that takes longer for smaller dust due to their longer settling timescale, approximately $\sim 1/(\tau_s \Omega)$ for $\tau_s \ll 1$ \citep{youdin_particle_2007}.

Finally, the length, time, and mass units used throughout this paper are $H,~ \Omega^{-1}$, and $\rho_{g0}H^3$, respectively. In our simulations, $H = \Omega = \rho_{g0} = 1$, allowing the results to be scaled to an arbitrary disk and location in it.

\begin{figure*}
    \centering
    \includegraphics[width=\textwidth]{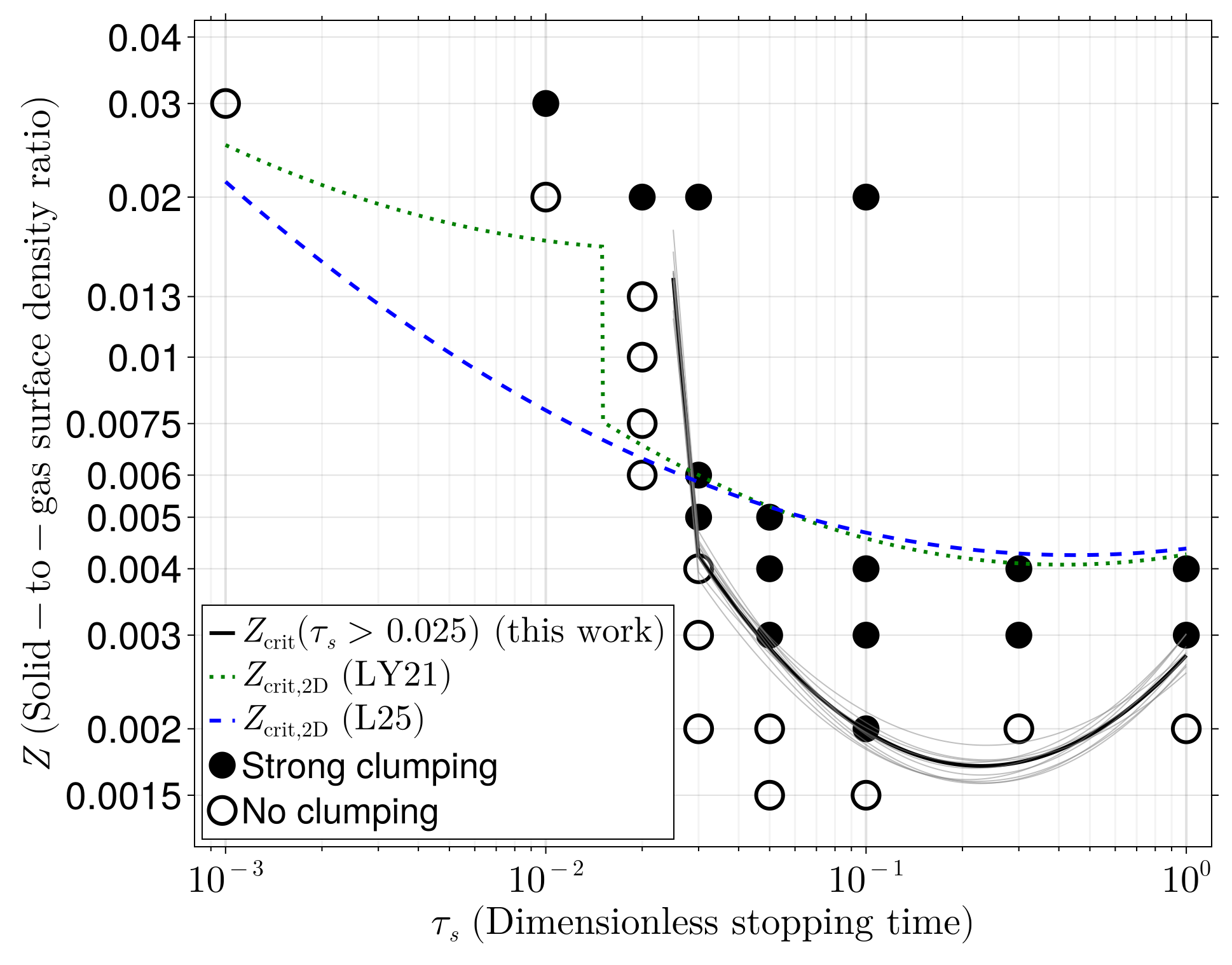}
    \caption{Overview of SI-driven clumping in 3D simulations with varying stopping times ($\tau_s$) and solid-to-gas surface density ratios ($Z$). In these runs, the global radial pressure gradient is set by $\Pi = 0.05$, and no external turbulence is included. Strong clumping occurs in runs marked by filled circles, while open circles indicate no clumping. The black curve is the best fit by least squares showing the minimum $Z$ value required for strong clumping as a function of $\tau_s$ ($\Zcrit$; see Equation \ref{eq:Zcrit} for the fitting parameters). Thin gray curves indicate the uncertainty range of the best fit, generated by randomly sampling the fitted parameters from their normal distributions. We provide the empirical fit only for $\tau_s > 0.025$ because of the insufficient number of data points for $\tau_s < 0.025$. Previous clumping thresholds from 2D axisymmetric simulations are shown as green, dotted \citepalias{LiYoudin21} and blue, dashed \citepalias{Lim25} curves. Critical $Z$ values range from $\sim 0.002$ at $\tau_s=0.1$ to above 0.03 at $\tau_s=10^{-3}$. Compared to 2D results, 3D simulations show lower $\Zcrit$ for $\tau_s > 0.02$, but higher $\Zcrit$ for $\tau_s < 0.02$.
    }
    \label{fig:Zcrit}
\end{figure*}

\begin{figure*}
    \centering
    \includegraphics[width=\textwidth]{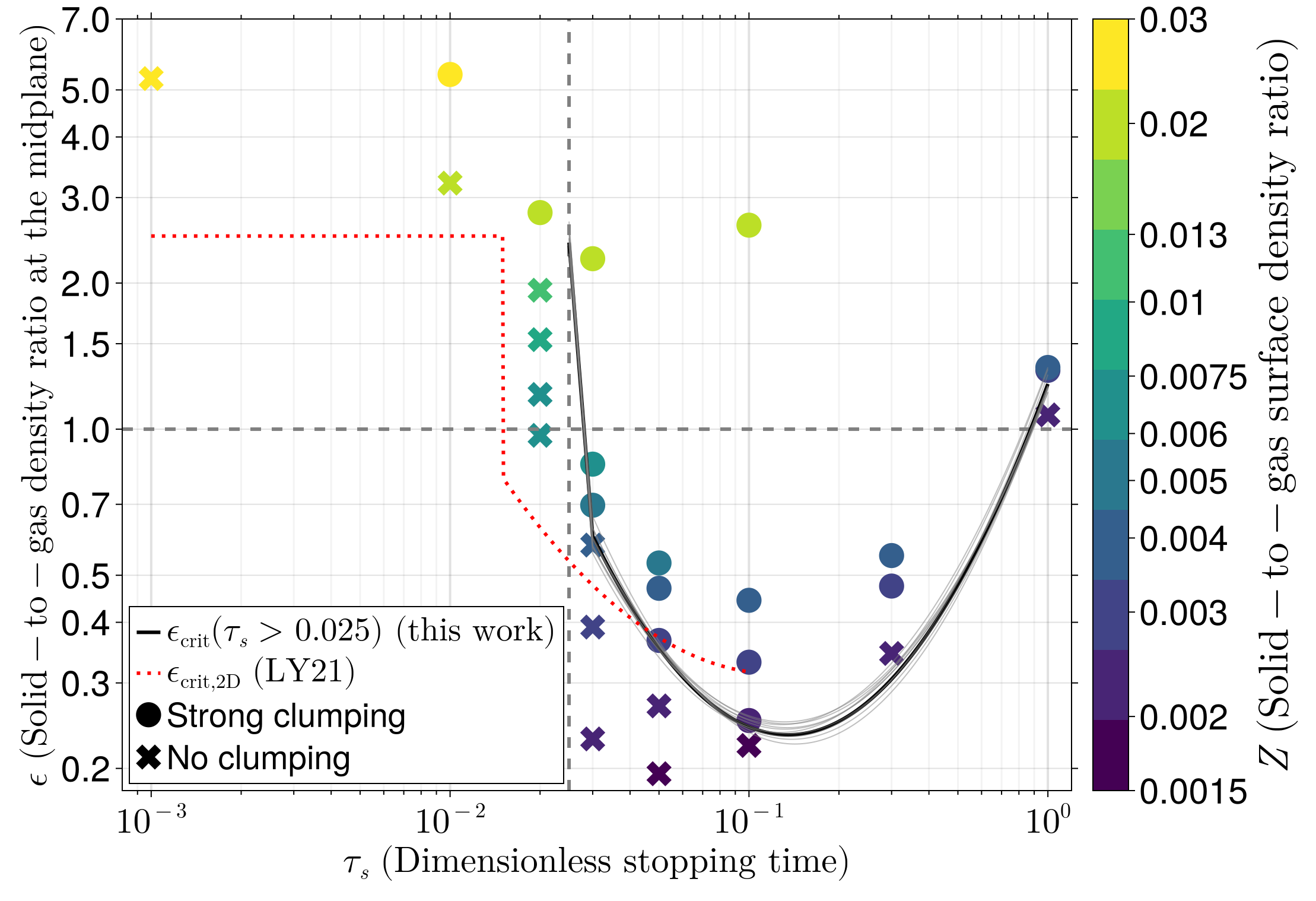}
    \caption{Midplane solid-to-gas density ratio ($\epsilon$) as functions of $\tau_s$ and $Z$, with the latter color-coded. Circles denote runs that show strong clumping, while x-markers indicate no clumping. The black curve represents the best fit for the critical midplane density ratio ($\epscrit$; see Equation \ref{eq:epscrit} for the fitting parameters) based on our 3D simulations. For the same reason as in $\Zcrit$ (Figure \ref{fig:Zcrit}), we truncate the fit at $\tau_s=0.025$. Thin gray curves exhibit the uncertainty range of the best fit. The red dotted curve shows $\epscrit$ determined from 2D simulations by \citetalias{LiYoudin21}. We truncate their curve at $\tau_s = 0.1$ due to a discrepancy in how $\epsilon$ is measured: we compute $\epsilon$ using $H_p^*$ (Equation~\ref{eq:Hpeff}), whereas they used $H_p$ (Equation~\ref{eq:Hp}). The difference between the two measurements is negligible for $\tau_s \leq 0.1$ but becomes significant for $\tau_s \geq 0.3$ (see Table \ref{table:simulations}). Based on our simulation results, for $\tau_s > 0.025$—where 3D simulations yield lower $\Zcrit$ than 2D—$\epscrit < 1$. By contrast, for $\tau_s < 0.025$—where 3D simulations yield higher $\Zcrit$—$\epscrit > 1$. Horizontal and vertical dashed lines mark $\epsilon = 1$ and $\tau_s = 0.025$, respectively, to illustrate this trend.}
    \label{fig:epscrit}
\end{figure*}

\section{Three-Dimensional Simulation Results} \label{sec:results_3D}
We perform a suite of 3D, vertically stratified simulations with different combinations of $\tau_s$ and $Z$, while fixing $\Pi = 0.05$. Our main goal is to determine the critical values of $Z$ ($\Zcrit$) and $\epsilon$ ($\epscrit$) above which strong clumping driven by the SI occurs across the range of $\tau_s$ values considered. A simulation is classified as a strong-clumping case if the maximum density of solid particles ($\rho_{p,\rm{max}}$) exceeds the Hill density, as defined in Equation \eqref{eq:rhoH}.

Table~\ref{table:simulations} lists time-averaged quantities related to solid particles for each simulation. The time averaging is performed over the pre-clumping phase \citepalias{LiYoudin21}, defined as the period between two time points: the time at which an equilibrium dust layer is established ($t_s$) and the time at which the maximum density of solids first exceeds $(2/3)\rho_H$ ($t_e$). For runs in which the maximum density never reaches the threshold, we set $t_e = t_{\rm{end}}$, i.e., the total simulation time.


As we will show below, both $\Zcrit$ and $\epscrit$ determined from our 3D simulations differ significantly from those in previous 2D, axisymmetric simulations. Moreover, we find that the evolution of dust filaments driven by the SI behaves differently in 2D vs 3D simulations. Given these substantial differences, we focus on the results from our 3D simulations in this section and present detailed comparisons with 2D simulations in Section \ref{sec:results:comp_2D}. 

\subsection{Critical Solid-to-gas Density Ratios for Strong Clumping}\label{sec:results_3D:crit_ratio}
\subsubsection{Critical $Z$}\label{sec:results:crit_ratio:Zcrit}
Figure~\ref{fig:Zcrit} shows conditions for strong clumping in terms of $\Zcrit(\tau_s)$. Each black circle represents a 3D simulation at a specific $(\tau_s,Z)$. Simulations that exhibit strong clumping are shown as filled circles, while those that do not are shown as open circles. The black solid curve denotes an empirical fit to $\Zcrit(\tau_s)$, above which strong clumping occurs; the fit’s uncertainty is indicated by thin gray curves. We describe the fitting procedure and provide the best-fit parameters below. For comparison, previously determined $\Zcrit(\tau_s)$ from 2D axisymmetric simulations are plotted as green, dotted \citepalias{LiYoudin21} and blue, dashed \citepalias{Lim25} curves.

In general, strong clumping occurs more readily—that is, at lower $Z$—for larger $\tau_s$. For example, $\Zcrit \sim 0.002-0.003$ for $\tau_s \geq 0.1$, while the critical value increases as $\tau_s$ decreases from 0.1, exceeding $\sim 0.03$ at $\tau_s = 10^{-3}$. This trend is consistent with previously determined thresholds from axisymmetric simulations \citepalias{carrera_how_2015,Yang2017,LiYoudin21,Lim25}.

The most striking feature of our $\Zcrit$ curve is its sharp rise as $\tau_s$ decreases from 0.03 to 0.02. A similar transition was previously reported in \citetalias{LiYoudin21}, where the jump occurred between $\tau_s = 0.01$ and 0.02 (see the green, dotted curve). Later, \citetalias{Lim25} conducted high-resolution axisymmetric simulations and revised the $\Zcrit$ value at $\tau_s = 0.01$ from above 0.01 to below, smoothing out the transition between $\tau_s = 0.02$ and 0.01 (see the blue, dashed curve). Despite this sharp jump in our 3D simulations, the reason for it is mostly physical and not numerical (see Section~\ref{sec:results:comp_2D:Ny} for details). However, resolution may be playing some minor role as well. We discuss the effect of grid resolution on our $\Zcrit$ curve in Section~\ref{sec:results:comp_2D:resolution}.


The black curve in Figure \ref{fig:Zcrit} is quadratic in logarithmic space, and fit
empirically by
\begin{equation}\label{eq:Zcrit}
	\begin{split}
    	\log{(\Zcrit)} = A(\log{\tau_s})^2 &+ B(\log{\tau_s}) + C  \\
    	& \quad \text{if } \tau_s > 0.025.
	\end{split}	
\end{equation}
To obtain the coefficients $A, B, C$, we perform a least squares fit assuming that, at a given $\tau_s$, $\Zcrit(\tau_s)$ lies between adjacent empty (no clumping) and filled (strong clumping) circles. According to the best fit, $A = 0.51,~B = 0.65,~C = -2.56$. The thin gray curves in Figure \ref{fig:Zcrit} are randomly drawn from a normal distribution of $(A, B, C)$ and thus represent the uncertainties in the best-fit curve. We provide the empirical fit to $\Zcrit$ only for $\tau_s > 0.025$ because only two $\tau_s$ values are available to fit the curve for $\tau_s < 0.025$, which is insufficient to reliably constrain the functional form of $\Zcrit$ in that regime.

\subsubsection{Critical $\epsilon$}\label{sec:results:crit_ratio:epscrit}
Figure \ref{fig:epscrit} presents $\epsilon$ values, calculated by Equation~\eqref{eq:eps}, for our fiducial 3D simulations with the critical curve ($\epscrit$; black curve) as a function of $\tau_s$. We color-code $Z$ values for each run, with the colorbar to the right of the plot. Strong-clumping and no-clumping runs are denoted by circle and `x' markers, respectively. The red, dotted curve is an empirical fit to $\epscrit(\tau_s)$ determined by 2D axisymmetric simulations of \citetalias{LiYoudin21}. We truncate their curve at $\tau_s = 0.1$ because of differences in how $\epsilon$ is estimated. Specifically, we calculate the effective scale height ($H_p^*$) using Equation~\eqref{eq:Hpeff}, and then estimate $\epsilon$ via Equation~\eqref{eq:eps}, while \citetalias{LiYoudin21} used $H_p$ (Equation~\ref{eq:Hp}) instead. The two approaches yield nearly identical values of $\epsilon$ for $\tau_s \leq 0.1$, but diverge significantly for $\tau_s \geq 0.3$ (see Table~\ref{table:simulations} to compare $H_p$ and $H_p^*$ in our simulations).

We obtain the best-fit curve for $\epscrit$ by applying the same fitting procedure used for $\Zcrit$. Assuming a quadratic form in logarithmic space, we fit
\begin{equation}\label{eq:epscrit}
	\begin{split}
    	\log{(\epscrit)} = A'(\log{\tau_s})^2 &+ B'(\log{\tau_s}) + C'  \\
    	& \quad \text{if } \tau_s > 0.025.
	\end{split}	
\end{equation}
with best-fit coefficients $A’ = 0.82$, $B’ = 1.24$, and $C’ = -0.2$. The thin gray curves in Figure \ref{fig:epscrit} represent the uncertainty in the fit, generated by randomly sampling the coefficients from a normal distribution of $(A',B',C')$. As with the $\Zcrit$ curve, we restrict the fit to $\tau_s > 0.025$ due to the limited number of data points at $\tau_s < 0.025$.

\begin{figure}[ht!]
    \includegraphics[width=\columnwidth]{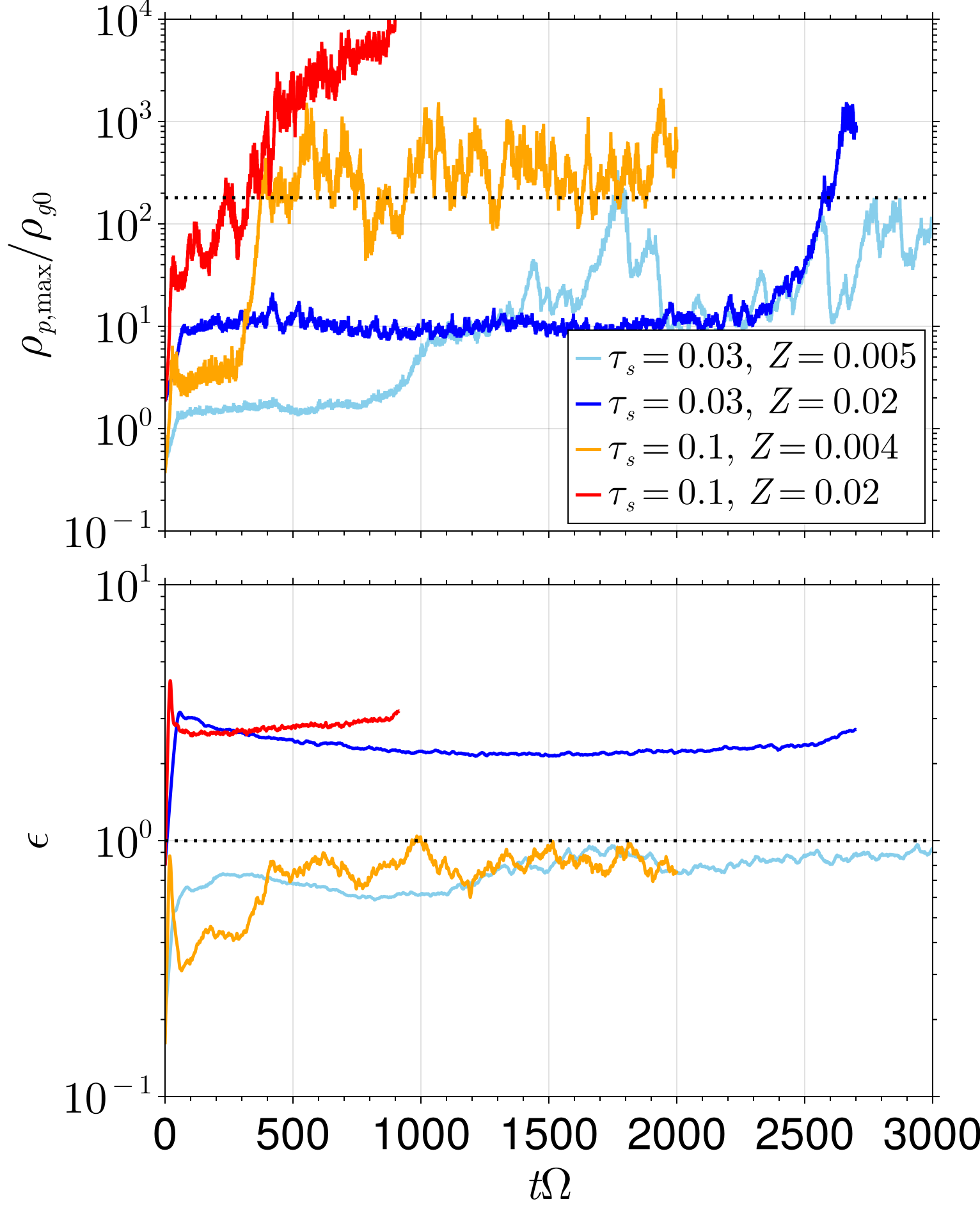}
    \centering 
    \caption{Time evolution of the maximum density of solid particles ($\rhopmax$; top) and the midplane density ratio ($\epsilon$; bottom) for strong-clumping runs with $\tau_s = 0.03$ and 0.1. For each $\tau_s$, two $Z$ values are selected to represent cases with $\epsilon < 1$ and $\epsilon > 1$. The horizontal lines mark the Hill density ($\rhoH$) in the top panel and $\epsilon = 1$ in the bottom panel. The runs with $\tau_s=0.1$ show similar evolutions of $\rhopmax$. By contrast, for $\tau_s = 0.03$, the lower-$Z$ case exhibits more stochastic evolution, whereas the higher-$Z$ case undergoes a longer period of quiescent evolution before $\rhopmax$ sharply rises toward the Hill density at $t\Omega \approx 2300$.}
    \label{fig:rhopmax-eps-tau003-tau01}
\end{figure}

We first describe our $\epscrit$ curve (black curve in Fig.~\ref{fig:epscrit}). As the curve shows, the critical $\epsilon$ values range from $\approx 0.25$ to $\approx 5$, with the minimum occuring at $\tau_s=0.1$. Similar to the $\Zcrit$ curve in Figure \ref{fig:Zcrit}, $\epscrit$ exhibits a sharp transition between $\tau_s=0.02$ and 0.03: as $\tau_s$ decreases from 0.03 to 0.02, the critical value increases from $\approx 0.6$ to $\approx 2.5$. As a result, $\epscrit < 1$ for $\tau_s > 0.025$ (to the right of the transition) and $\epscrit > 1$ for $\tau_s < 0.025$ (to the left of the transition).

We now compare our results with those of \citetalias{LiYoudin21} (red, dotted curve). At $\tau_s = 0.03$, our critical $\epsilon$ values closely match theirs and become even lower as $\tau_s$ increases toward 0.1. By contrast, for $\tau_s \leq 0.02$, our critical $\epsilon$ values are higher by a factor of a few. Interestingly, this divergence aligns with where $\epscrit$ crosses unity: $\epscrit < 1$ for $\tau_s \geq 0.03$, and within this regime, our 3D simulations yield lower $\Zcrit$ than previous 2D simulations (see Fig.~\ref{fig:Zcrit}). Conversely, for $\tau_s \leq 0.02$, where $\epsilon_{\rm crit} > 1$, our 3D simulations produce higher $\Zcrit$ than the 2D results. In other words, based on Figures~\ref{fig:Zcrit}–\ref{fig:epscrit}, strong clumping occurs more easily in 3D simulations than in 2D when $\epscrit < 1$, but is more difficult when $\epscrit > 1$. Along these lines, we present distinct clumping mechanisms driven by the SI in the $\epsilon < 1$ and $\epsilon > 1$ regimes in the next subsection. 

\begin{figure*}
    \centering
    \includegraphics[width=\textwidth]{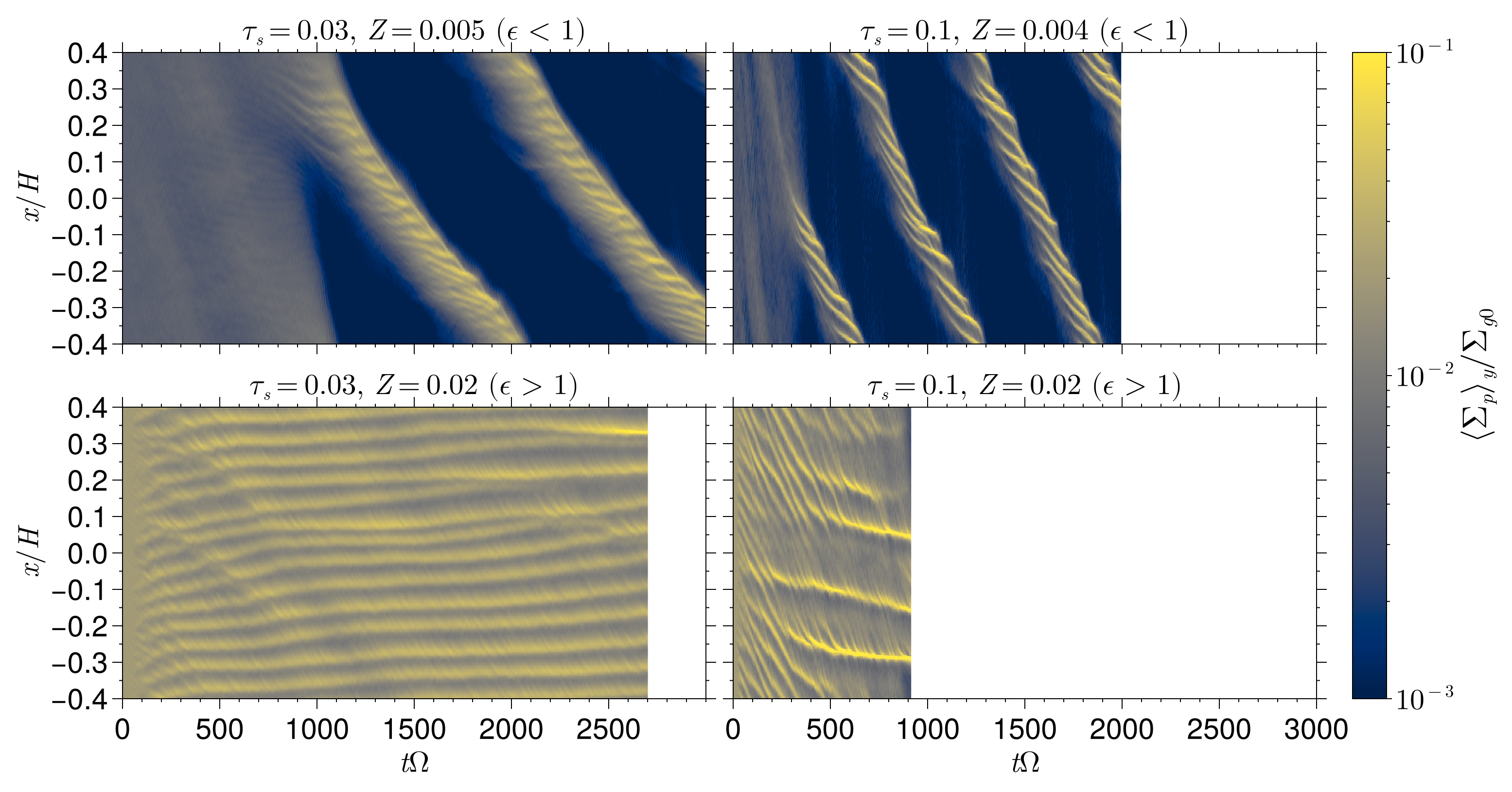}
    \caption{Azimuthally averaged surface density of solid particles $\langle \Sigma_p \rangle_y$ vs $x$ and $t$, illustrating the evolution of filaments formed by the SI. Left and right columns correspond to $\tau_s=0.03$ and 0.1 cases, respectively, while upper and lower panels represent cases with $\epsilon < 1$ and  $\epsilon > 1$ at each $\tau_s$. A clear morphological bifurcation is observed between the two regimes: the $\epsilon < 1$ cases exhibit a filaments-in-filaments structure, whereas the $\epsilon > 1$ cases develop multiple, evenly spaced filaments (until they merge) across the radial domain.}
    \label{fig:rhopt-tau003-tau01}
\end{figure*}

\subsection{How $\epsilon$ Regulates the Formation of Dust Filaments}\label{sec:results_3D:SI_clumping}

In this subsection, we examine how $\epsilon$ influences the formation and evolution of dust filaments. We focus on two values of $\tau_s$: 0.03 and 0.1. For each $\tau_s$, we consider two $Z$ values that represent the $\epsilon < 1$ and $\epsilon > 1$ regimes. Specifically, we use $Z = 0.005$ and 0.004 to represent $\epsilon < 1$ cases for $\tau_s=0.03$ and 0.1, respectively. For $\epsilon > 1$ cases, we use $Z=0.02$ for both values of $\tau_s$. All of the selected simulations show strong clumping. 

Figure~\ref{fig:rhopmax-eps-tau003-tau01} shows the time evolution of the maximum density of solid particles ($\rhopmax$; upper panel) and the midplane density ratio $\epsilon$ (lower panel) for the four selected simulations. The runs with $\tau_s = 0.03$ are shown in lightblue ($Z = 0.005$) and blue ($Z = 0.02$), while those with $\tau_s = 0.1$ are shown in orange ($Z = 0.004$) and red ($Z = 0.02$). As indicated in the lower panel, the $Z = 0.004$–0.005 runs correspond to $\epsilon < 1$, whereas the $Z = 0.02$ runs correspond to $\epsilon > 1$. The horizontal lines mark the Hill density ($\rhoH$; Equation~\ref{eq:rhoH}) in the upper panel and $\epsilon = 1$ in the lower panel.

We first compare different $Z$ values at $\tau_s = 0.03$.  The $Z = 0.02$ run maintains a long saturated state between $t\Omega \approx 100$ and $\approx 2300$ with $\rhopmax \approx 10 \rho_{g0}$, after which $\rhopmax$ rises sharply and surpasses the Hill density. On the other hand, the $Z = 0.005$ run exhibits two distinct saturated phases: the first between $t\Omega \approx 100$ and 800 with $\rhopmax \sim \rho_{g0}$, and the second from $t\Omega \approx 1100$ to the end of the simulation, characterized by more stochastic evolution and higher values of $\rhopmax$ than those during the first saturated phase. This distinguishing behavior is closely tied to the evolution of filaments. 

Figure~\ref{fig:rhopt-tau003-tau01} present spacetime plots of the vertically integrated and azimuthally averaged density of solids, $\langle \Sigma_p \rangle_y$, as functions of time and radial position $x$. The panels on the left correspond to $\tau_s = 0.03$. Comparing the upper ($Z = 0.005$) and lower ($Z = 0.02$) left panels, the filament evolution differs markedly with $Z$.

In the run with $Z = 0.005$, no dense filaments are present until $t\Omega \approx 800$, corresponding to the first saturated state (see the light blue curve in the upper panel of Fig.~\ref{fig:rhopmax-eps-tau003-tau01}). Filaments begin to emerge  around $t\Omega \approx 800$, accompanied by an increase in $\rhopmax$. The second saturated state, between $t\Omega \approx 1100$ and 3000, is characterized by a filaments-in-filaments structure—first identified in \citetalias{LiYoudin21}—in which subfilaments form within a high density region (the parent filament). The parent filament is persistent, while the subfilaments within it reach higher density and occasionally merge, contributing to the stochastic evolution of the maximum density and driving it toward and beyond the Hill density.

By contrast, the $Z = 0.02$ run exhibits multiple filaments as early as $t\Omega \approx 100$, with a greater number compared to the $Z = 0.005$ case (see the lower left panel of Fig.~\ref{fig:rhopt-tau003-tau01}). These filaments are regularly spaced and remain nearly stationary in radial position; however, this reflects the motion of the pattern, which is decoupled from the motion of the material (i.e., the particles) themselves. The quiescent evolution of these filaments corresponds to the saturated state of $\rhopmax$ seen in the $Z = 0.02$ run (blue curve in Fig.~\ref{fig:rhopmax-eps-tau003-tau01}) before the  increase toward the Hill density. This late-stage increase is consistent with the filament located at $x \sim 0.3H$ (lower left panel of Fig.~\ref{fig:rhopt-tau003-tau01}) gradually growing in density as solids from upstream join the filament.

\begin{figure*}[ht!]
	\includegraphics[width=\textwidth]{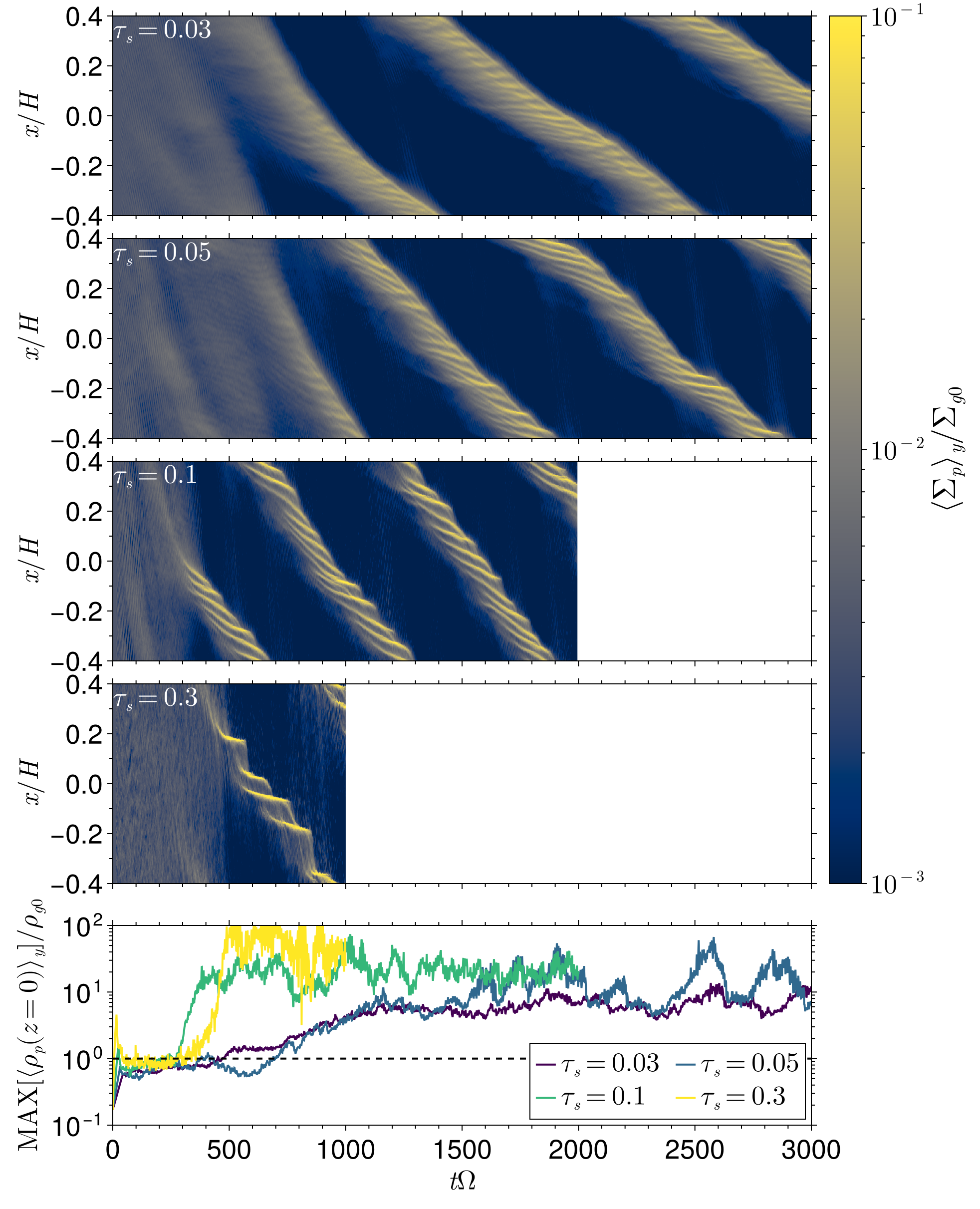}\
	\caption{Spacetime plots for different values of $\tau_s$—0.03, 0.05, 0.1, and 0.3—are shown from the top to the second-to-last panels, all with $Z = 0.004$. The bottom panel shows the time evolution of the maximum midplane solid-to-gas density ratio (which is not equal to $\epsilon$ as defined in Equation~\ref{eq:eps}). All simulations have $\epsilon$ less than 1 and exhibit a filaments-in-filaments s structure. The emergence of this structure roughly coincides with the point at which the maximum midplane density ratio begins to exceed unity, as indicated by the horizontal line in the bottom panel.
    }
	\label{fig:rhopt-Z0004}
\end{figure*} 
We now focus on the runs with $\tau_s = 0.1$. As shown in Figure~\ref{fig:rhopmax-eps-tau003-tau01}, the maximum density in both runs rapidly reaches the Hill density around $t\Omega \approx 300$. In the run with $Z = 0.004$ (orange), $\rhopmax$ remains in a saturated state between $t\Omega \approx 100$ and 300 before rising sharply. By contrast, $Z = 0.02$ case (red) exhibits a continuous increase without a saturated state, eventually reaching $\sim 10^4 \rho_{g0}$ by the end of the simulation.

The behavior of SI-driven filaments at $\tau_s = 0.1$ is similar to those at $\tau_s = 0.03$ in terms of their dependence on $\epsilon$. The right panels of Figure~\ref{fig:rhopt-tau003-tau01} show this: a filaments-in-filaments structure forms when $Z = 0.004$ ($\epsilon < 1$), whereas multiple filaments develop when $Z = 0.02$ ($\epsilon > 1$). In the case of $Z = 0.004$, the saturated phase between $t\Omega=100$ and 300 corresponds to the development of the filaments-in-filaments structure, with the sharp rise in $\rhopmax$ coinciding with its emergence. When $Z = 0.02$, multiple filaments form early across the radial domain, similar to the corresponding run at $\tau_s = 0.03$. However, filament merging is more prominent at $\tau_s = 0.1$. These frequent mergers lead to more stochastic evolution of $\rhopmax$ and higher values compared to the $\tau_s = 0.03$ case. This behavior can be attributed to the faster radial drift of larger solids (or, larger $\tau_s$) at a given $\rho_p/\rho_g$, as predicted by the equilibrium radial velocity solution \citep{Nakagawa1986,YG05}.

\begin{figure*}[ht!]
	\centering
	\includegraphics[width=\textwidth]{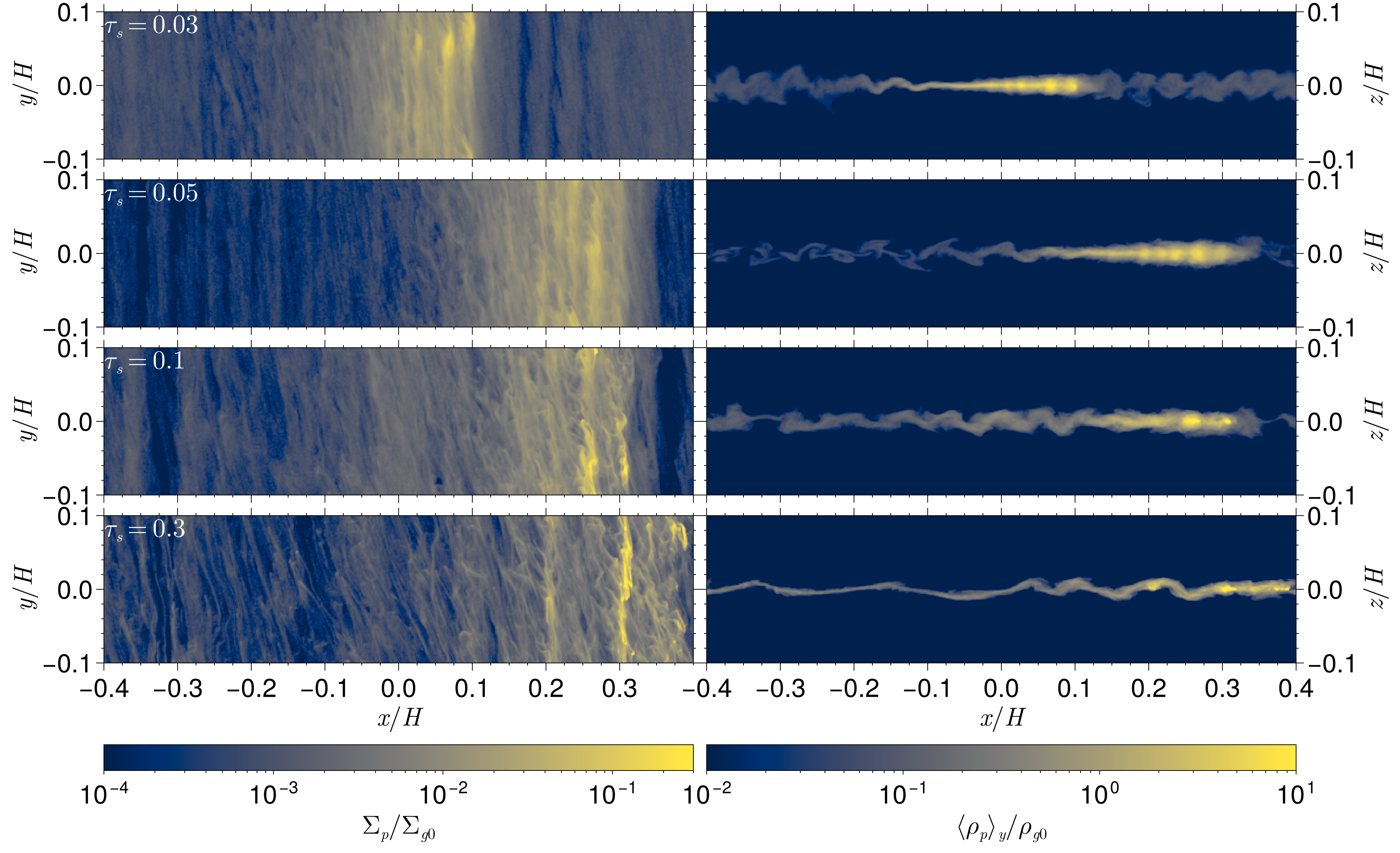}
	\caption{Final snapshots of particle density for different values of $\tau_s$—0.03, 0.05, 0.1, and 0.3 from top to bottom—with a fixed $Z$ value of 0.004 (the same runs as in Fig.~\ref{fig:rhopt-Z0004}). The left panels show the surface density $\Sigma_p$ (top view), while the right panels show the azimuthally averaged density $\langle \rho_p \rangle_y$ (side view). All simulations have $\epsilon < 1$ during the pre-clumping phase. Small-scale non-axisymmetric clumping is clearly visible in the left panels. The structure of the vertical layer varies both along the radial direction within each simulation and with $\tau_s$ across simulations as can be seen from the right panels.}
	\label{fig:snapshots-eps-bw-1}
\end{figure*}


In summary, filament formation driven by the SI manifests as a filaments-in-filaments structure in our simulations with $\epsilon < 1$. In fact, as we show in Figure~\ref{fig:rhopt-Z0004}, we find that the filaments-in-filaments structure appears for $0.03\leq \tau_s\leq 0.3$, the range over which most of our simulations yield $\epsilon < 1$ (see Fig.~\ref{fig:epscrit}). The simulations shown in the figure have the same $Z$ value of 0.004. The bottom panel displays the time evolution of the maximum value of $\langle \rho_p(z=0) \rangle_y$ in units of $\rho_{g0}$. Comparing the spacetime plots to the evolution of the maximum ratio, we find that the filaments-in-filaments structure tends to emerge around the time when this ratio exceeds unity. This suggests that with a small mass budget, solid particles gradually concentrate in the radial direction, increasing the local midplane density ratio to the point where a filaments-in-filaments structure emerges.

By contrast, when $\epsilon > 1$, the available solid mass is sufficient to rapidly form multiple filaments. Rather than being locally concentrated, these filaments tend to be evenly spaced. This configuration leads to either rapid filament merging—as in the $\tau_s = 0.1$, $Z = 0.02$ case—or more quiescent filament evolution, as seen in the case of $\tau_s = 0.03$, $Z = 0.02$, depending on the drift speed set by $\tau_s$. Somewhat unexpectedly, the more quiescent filament behavior at $\tau_s = 0.03$ delays clumping, taking longer to reach the Hill density than the $Z = 0.005$ case at the same $\tau_s$.

We caution that the number of filaments formed also depends on $\tau_s$. For example, in our simulations with $\tau_s = 10^{-3}$ and $10^{-2}$ (both with $Z = 0.03$), $\epsilon > 1$ is achieved (see Fig.~\ref{fig:epscrit}), yet no filaments form at $\tau_s=10^{-3}$, and only a few appear at $\tau_s=10^{-2}$. While the smaller radial domain size ($L_x = 0.4H$ instead of $0.8H$) used in these runs may contribute to this outcome, it is well established that the SI becomes less efficient at forming filaments—and that $\Zcrit$ increases—as $\tau_s$ decreases \citepalias{carrera_how_2015,Yang2017, LiYoudin21, Lim25}. 

Finally, we find that the filaments-in-filaments structure is a non-axisymmetric phenomenon that does not appear in 2D axisymmetric simulations. We return to this point in Section~\ref{sec:results:comp_2D}, where we compare results from 2D and 3D simulations. Additionally, we note that in all simulations exhibiting the filaments-in-filaments  structure, only a single parent filament is observed. As a result, it remains uncertain how large the radial domain ($L_x$) must be to allow the formation of multiple parent filaments.

\subsection{Structures of Dust Layer}\label{sec:results_3D:dust_layer}
Under vertical gravity, solids undergo sedimentation toward the disk midplane, counteracted by vertical diffusion driven by gas. When these two effects balance, an equilibrium vertical layer forms. In this section, we examine how the structure of the dust layer—both in the $x$–$y$ and $x$–$z$ planes—varies with $\tau_s$ and $\epsilon$ in our 3D simulations.  

Figure \ref{fig:snapshots-eps-bw-1} shows final snapshots of $\rho_p$ for different $\tau_s$ values but with the same $Z$ value of 0.004. From top to bottom, the panels correspond to $\tau_s=0.03,~0.05,~0.1$, and 0.3, respectively. Left and right panels show surface density of solids, $\Sigma_p$, and azimuthally averaged density, $\langle \rho_p \rangle_y$, respectively. 

Small-scale non-axisymmetric concentrations are evident across all $\tau_s$ values, as shown in the left panels. These clumps are embedded within azimuthally elongated dense filaments and attain the highest $\rho_p$ values in regions near the midplane. In addition, these dense filaments are closely spaced in the radial direction, consistent with the filaments-in-filaments structures observed in Figure~\ref{fig:rhopt-Z0004}. 

The right panels reveal that the structure of the vertical layer varies with both local density and $\tau_s$. For $\tau_s = 0.03$ and 0.05, the vertical layer appears slightly thinner in high-density regions compared to low-density regions. In the $\tau_s = 0.1$ case, this dependence is weaker, with the layer thickness remaining more uniform. At $\tau_s = 0.3$, the vertical structure displays a wave-like pattern—also reported in 2D simulations (e.g., \citetalias{LiYoudin21})—and appears slightly thicker in regions of higher density. 

We find that the wave-like pattern is most pronounced when $\tau_s=0.3$ and 1.0 (not shown), which is the case in 2D simulations as well \citepalias{LiYoudin21}. This undulated layer contributes to the difference between $H_p$ and $H_p^*$. This is because while $H_p$ measures the dispersion of the vertical position of particles around the midplane—as $\overline{z_p}$ in Equation~\eqref{eq:Hp} is nearly zero, $H_p^*$ measures the dispersion relative to the center of a dust layer at a given column. As a result, $H_p \simeq 2 \langle H_p^* \rangle$ for $\tau_s = 0.3$ and 1.0, where the wave-like structure is prominent, whereas $H_p \simeq \langle H_p^* \rangle$ for $10^{-3} \leq \tau_s \leq 10^{-1}$, where such vertical undulations are absent.  

\begin{figure}
	\centering
	\includegraphics[width=\columnwidth]{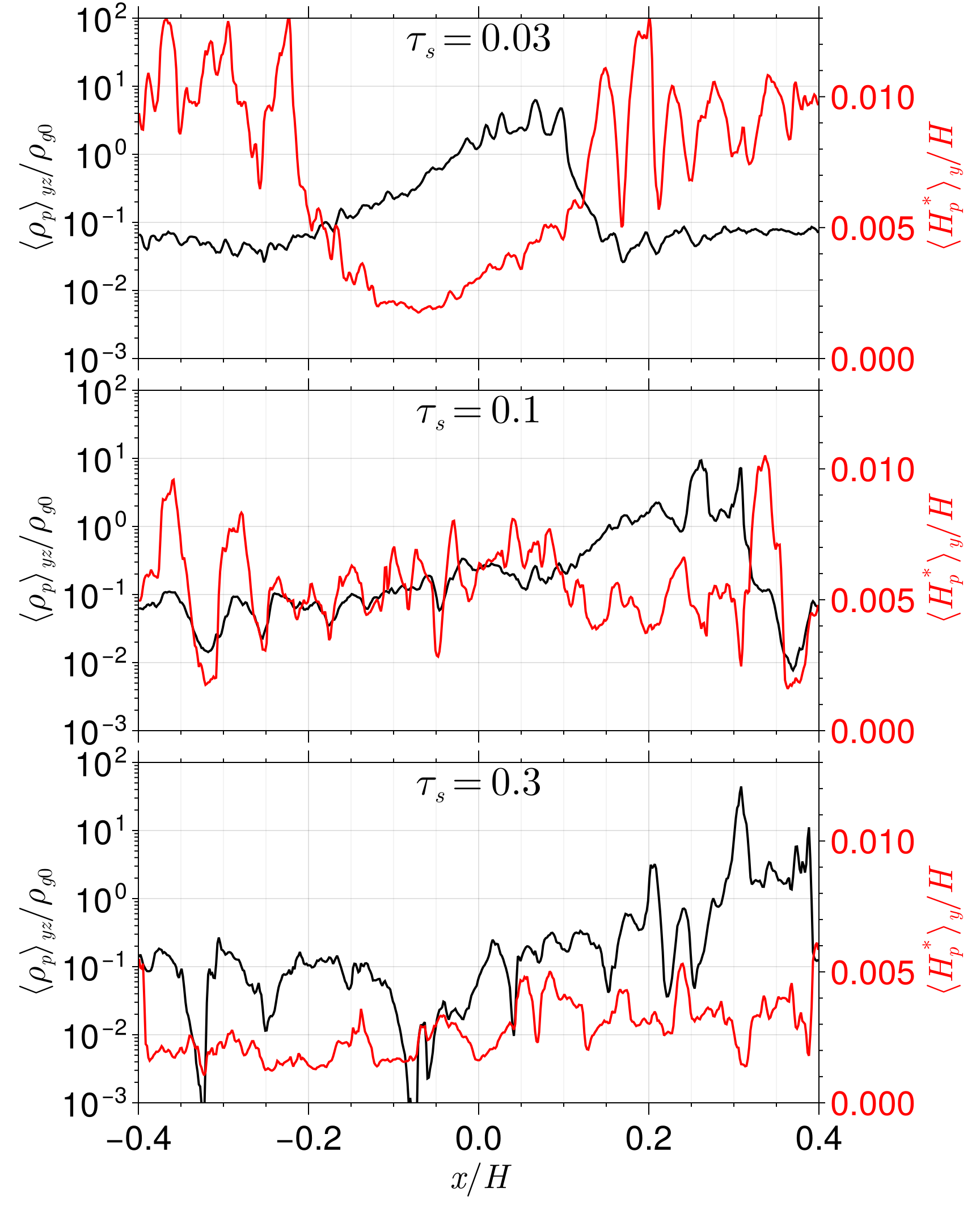}
	\caption{Final snapshots of radial profiles of solid density averaged over the $y$–$z$ plane, $\langle \rho_p \rangle_{yz}$ (black; left axis), and azimuthally averaged scale height of solids, $\langle H_p^* \rangle_y$ (red; right axis), for the runs shown in Figure~\ref{fig:snapshots-eps-bw-1}, excluding the $\tau_s = 0.05$ case. The vertical averaging of $\rho_p$ is performed within $\pm \langle H_p^* \rangle$ around the midplane. In the runs with $\tau_s = 0.03$ and 0.1, $H_p^*$ is smaller in high-density regions compared to low-density regions, with the reduction more pronounced in the former. By contrast, for $\tau_s = 0.3$, $H_p^*$ is slightly larger in high-density regions.}
	\label{fig:rhop-Hpeff}
\end{figure}

To examine the vertical layer structure in greater detail, Figure \ref{fig:rhop-Hpeff} presents radial profiles of azimuthally and vertically averaged density of solids, $\langle \rho_p \rangle_{yz}$ (left axis), and azimuthally averaged scale height, $\langle H_p^* \rangle_y$ (right axis; see Equation \ref{eq:Hpeff}).  The density is averaged over the full azimuthal extent and within $\pm \langle H_p^* \rangle$ in the vertical direction. From top to bottom, the panels show the profiles of $\tau_s=0.03,~0.1$, and 0.3, respectively. 

When $\tau_s = 0.03$, $H_p^*$ is reduced in high-density regions where filaments reside ($x/H \approx -0.1$ to 0.1), suggesting that increased inertia of solids suppresses turbulent stirring, consistent with the simulations of \cite{Lim24}. However, $H_p^*$ increases with increasing particle density within $x/H \approx -0.1$ to 0.1. This is also seen in the upper right panel of Figure~\ref{fig:snapshots-eps-bw-1} where the dust layer is extremely thin around $x/H = -0.1$ and gradually thickens toward $x/H \approx 0.1$, approaching filaments. We do not fully understand why within certain high density regions, $H_p^*$ decreases and yet increases in others. It seems likely that there is some critical density above which the suppression of turbulence overpowers the stirring of it. However, delving into this in more detail will need to be addressed in future work.

The run with $\tau_s = 0.1$ shows similar trends, though not to the same extent as in the case of $\tau_s=0.03$. While $H_p^*$ is reduced in high-density regions ($x/H \in [0.1, 0.3]$), the reduction is less pronounced than in the $\tau_s = 0.03$ case. Moreover, $H_p^*$ is nearly independent of $\rho_p$ within the dense regions. Overall, the distribution of $H_p^*$ is relatively uniform across the radial domain.

For $\tau_s = 0.3$, the radial profile of $H_p^*$ also appears relatively uniform, but its dependence on $\rho_p$ reverses compared to the $\tau_s = 0.03$ case. That is, $H_p^*$ is slightly larger in high-density regions than in low-density regions, implying that increased inertia of solids enhances stirring rather than suppressing it. Conversely, the smaller $H_p^*$ values in low-density regions suggest that weaker backreaction results in more efficient settling rather than increased diffusion—opposite to the behavior seen in the $\tau_s = 0.03$ case.

Overall, the results from Figures~\ref{fig:snapshots-eps-bw-1}–\ref{fig:rhop-Hpeff} suggest that $\partial H_p / \partial \rho_p < 0$ for $\tau_s < 0.1$, meaning that more solids reduce stirring, whereas $\partial H_p / \partial \rho_p > 0$ for $\tau_s > 0.1$, meaning that more solids enhance stirring. This derivative seems to be approximately zero when $\tau_s = 0.1$. This finding is consistent with \citetalias{LiYoudin21}, who measured $H_p$ in their 2D simulations over the range $10^{-3} \leq \tau_s \leq 1.0$. They found that $\partial H_p / \partial Z < 0$ for $\tau_s < 0.1$ and $\partial H_p / \partial Z > 0$ for $\tau_s > 0.1$ (see their Fig.~4).  While their analysis focused on the dependence of $H_p$ on the global solid abundance $Z$, rather than on local values of $\rho_p$ as we examine here, our results show qualitative agreement with theirs.
\begin{figure*}[ht!]
	\centering
	\includegraphics[width=\textwidth]{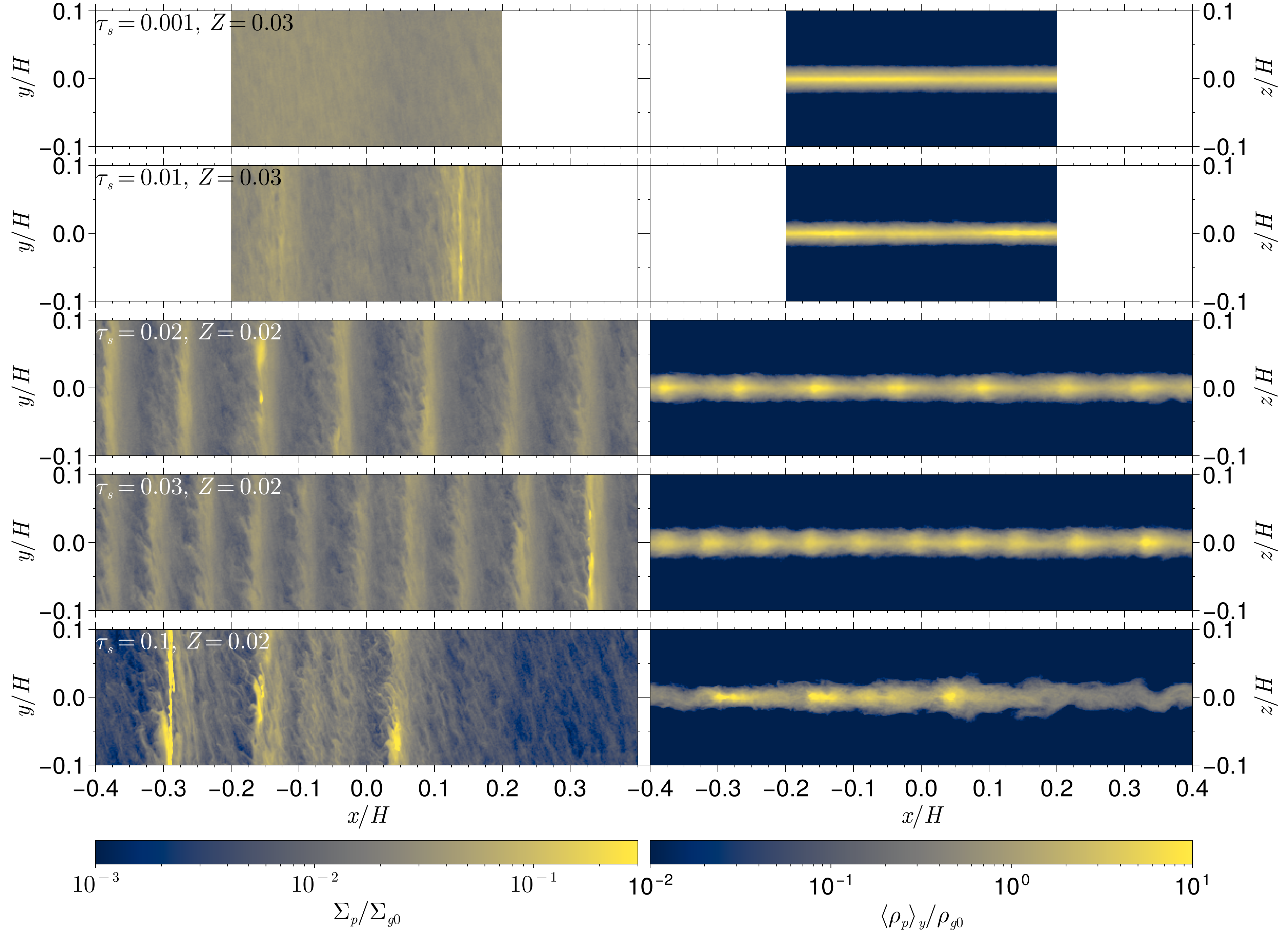}
	\caption{Same as Figure~\ref{fig:snapshots-eps-bw-1}, but for simulations with $\epsilon > 1$ during the pre-clumping phase. The values of $\tau_s$ and $Z$ are shown in the upper left corner in left panels. As in the $\epsilon < 1$ cases shown in Figure~\ref{fig:snapshots-eps-bw-1}, non-axisymmetric concentration is evident across all simulations except for $\tau_s=0.001$ (left panels). However, the structure of the vertical layer exhibits weaker dependence on the radial position ($x$) within each simulation and on $\tau_s$ across simulations (right panels), compared to the $\epsilon < 1$ cases.}
	\label{fig:snapshots-eps-ab-1}
\end{figure*}


\begin{figure*}[ht!]
	\centering
	\includegraphics[width=\textwidth]{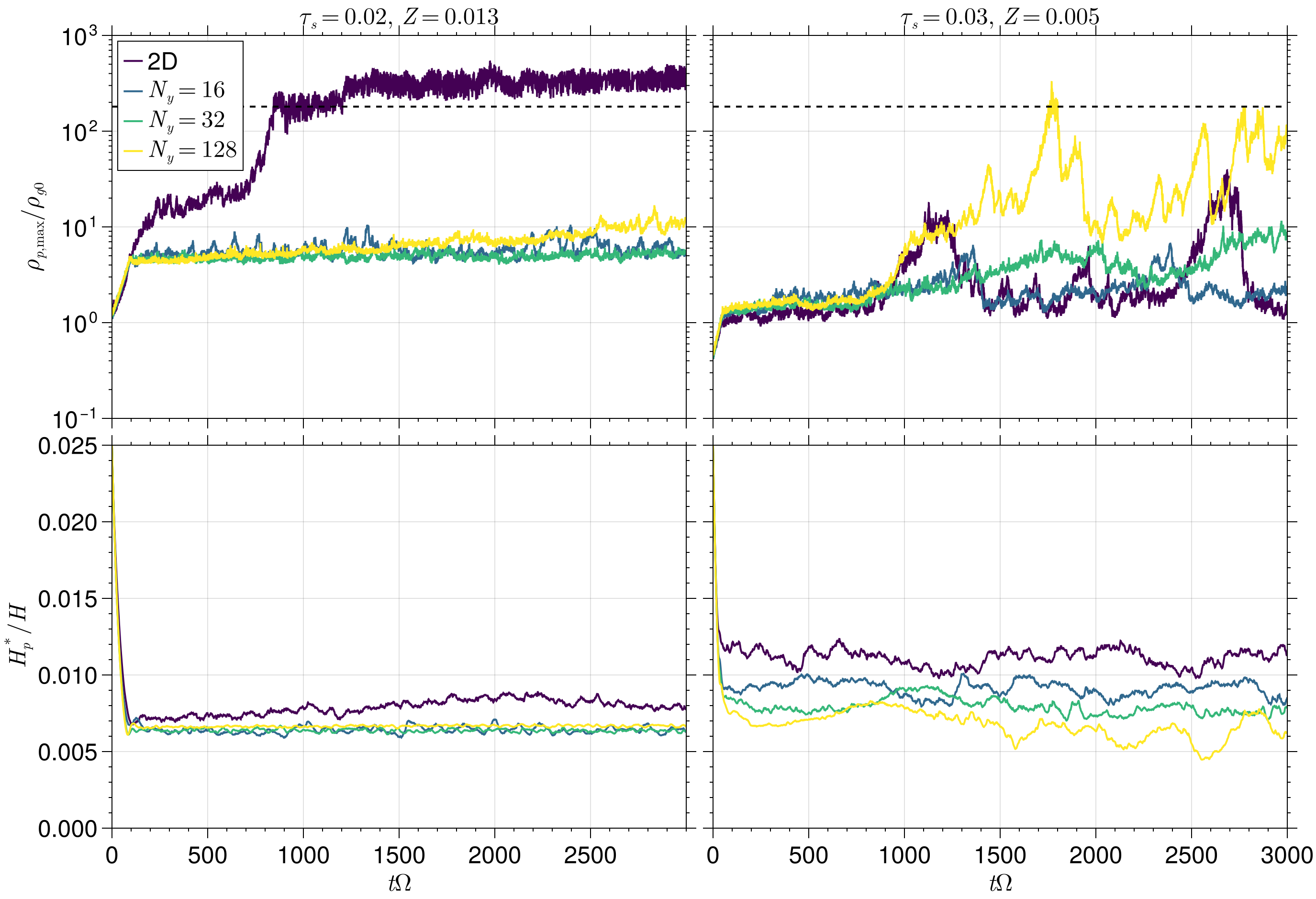}
	\caption{Maximum  density of solids (upper panels) and their scale height $H_p^*$ (lower panels) as functions of time. The left and right columns correspond to $\tau_s = 0.02$, $Z = 0.013$ ($\epsilon > 1$) and $\tau_s = 0.03$, $Z = 0.005$ ($\epsilon < 1$), respectively. For each parameter set, simulations with varying numbers of grid cells in the azimuthal ($N_y$) direction are shown: $N_y = 1$ (2D), 16, 32, and 128 (fiducial 3D model). The horizontal line in the upper panels marks the Hill density, as defined in Equation~\eqref{eq:rhoH}. For $\tau_s = 0.02$, $Z = 0.013$, only the 2D model shows strong clumping, while for $\tau_s = 0.03$, $Z = 0.005$, strong clumping occurs only in the fiducial 3D run. In both cases, $H_p^*$ is largest in the 2D simulations.
		}
		\label{fig:rhopmax-Hp-tau002-tau003-Ny}
\end{figure*}

Figure~\ref{fig:snapshots-eps-ab-1} is similar to Figure~\ref{fig:snapshots-eps-bw-1}, but for cases with $\epsilon > 1$ and smaller values of $\tau_s$. As in the $\epsilon < 1$ cases, the surface density distribution (left panels) exhibits non-axisymmetric concentration except for $\tau_s=0.001$. However, unlike in the $\epsilon < 1$ regime, the filaments are evenly spaced—except in the $\tau_s = 0.1$ case, where filament merging occurs over time (see Fig.~\ref{fig:rhopt-tau003-tau01}).

The structure of the vertical layer does not vary significantly with $\rho_p$ in each simulation compared to the $\epsilon < 1$ cases. For example, the run with $\tau_s = 0.03$ shows little variation in vertical layer thickness along the radial direction, whereas the run with the same $\tau_s$ but with $Z = 0.004$ (top panels in Fig.~\ref{fig:snapshots-eps-bw-1}) exhibits substantial variation. This contrast arises because, with $\epsilon > 1$, $\rho_p$ varies little along the radial direction resulting from the quiescent evolution of filaments (see Fig.~\ref{fig:rhopt-tau003-tau01}). Since $H_p^*$ is sensitive to local density, its spatial variation is also vanishingly small. However, we caution that $\tau_s$ also contributes to the variation of $\rho_p$ by increasing drift speeds of filaments and thus promoting filaments merging as $\tau_s$ becomes larger, as in the case of $\tau_s=0.1$.  

On top of that, the vertical layer structure does not vary significantly with $\tau_s$. This is not the case in previous 2D simulations, where strong dependence on $\tau_s$ is seen  \citepalias{LiYoudin21}. For example, they found a striking difference in vertical dust layer structure between $\tau_s = 10^{-3}$ and $10^{-2}$. Specifically, the $\tau_s = 10^{-2}$ case exhibits a feathered surface, whereas the $\tau_s = 10^{-3}$ shows a fish-shaped structure with a strongly corrugated surface (see their Fig.~5).

Moreover, when comparing our 3D simulations at $\tau_s = 10^{-3}-10^{-2}$ to their 2D counterparts, we find that the vertical layers in our models exhibit significantly less surface corrugation (see also \citetalias{Yang2017} for a similar finding). This morphological difference is most pronounced at $\tau_s = 10^{-3}$—compare the top right panel of Figure~\ref{fig:snapshots-eps-ab-1} with Figure 5 of \citetalias{LiYoudin21}. As a result, our $\tau_s = 10^{-3}$, $Z = 0.03$ run yields a smaller dust scale height ($0.006H$ vs. $0.009H$) and thus a higher midplane dust-to-gas ratio ($\epsilon = 5.2$ in 3D vs. $\epsilon = 3.5$ in 2D). Despite the higher $\epsilon$, strong clumping does not occur in our 3D model, whereas it does in their 2D simulation. This discrepancy motivates a detailed comparison between 2D and 3D simulations, which we present in the next section.

\section{Comparison to Axisymmetric Simulations}\label{sec:results:comp_2D}
One of the key findings from the previous section is that $\Zcrit$ determined by our 3D simulations exhibits a sharp jump as $\tau_s$ decreases from 0.03 to 0.02 (Fig.~\ref{fig:Zcrit}). Furthermore, a clear trend appears when comparing our results with previous 2D simulations: for cases with $\epsilon < 1$, our 3D simulations yield lower $\Zcrit$ values, while for $\epsilon > 1$, they result in higher $\Zcrit$. This discrepancy between 2D and 3D results occurs near the transition point where $\Zcrit$ sharply increases (Fig.~\ref{fig:epscrit}).

In this section, we systematically compare SI-driven clumping between 2D and 3D simulations using two approaches. In Section~\ref{sec:results:comp_2D:Ny}, we examine the effect of the azimuthal dimension at our fiducial grid resolution of $640/H$. To this end, we select two representative parameter sets: $\tau_s = 0.02$, $Z = 0.013$ and $\tau_s = 0.03$, $Z = 0.005$. The former lies to the left of the sharp jump in $\Zcrit$ and is below the threshold, with $\epsilon > 1$, while the latter lies to the right of the jump and is above the threshold, with $\epsilon < 1$. For each case, we run simulations with four different values of $N_y$ (the number of grids in $y$ directions): 1, 16, 32, and 128, where $N_y = 1$ corresponds to a 2D axisymmetric model, and $N_y = 128$ is our fiducial 3D setup. This allows us to assess how the azimuthal dimension influences SI clumping. We use smaller azimuthal domain sizes ($L_y$) in 3D runs (i.e., $N_y>1$) with smaller $N_y$ to maintain a fixed grid resolution of $640/H$.  Specifically, $L_y/H = 0.025$, 0.05, and 0.2 correspond to $N_y = 16$, 32, and 128, respectively. Throughout this section, we use $N_y$ to refer to simulations with different azimuthal extents. 

On top of that, since the efficiency of SI-driven clumping is known to be sensitive to grid resolution \citepalias{Yang2017,LiYoudin21,Lim25}, we also conduct a resolution test at $\tau_s = 0.02$, $Z = 0.013$ for both 2D and 3D models and present the result in Section~\ref{sec:results:comp_2D:resolution}. This test aims to examine how the effect of grid resolution on SI clumping differs between 2D and 3D and to check whether increasing resolution lowers our $\Zcrit$ at this $\tau_s$, potentially smoothing the sharp transition; this latter test is strongly motivated by our 2D simulations in \citetalias{Lim25}, which showed that higher resolution can remove the discontinuity found in \citetalias{LiYoudin21}, as we explained in Section~\ref{sec:results:crit_ratio:Zcrit}.

\subsection{Effect of Azimuthal Domain Size}\label{sec:results:comp_2D:Ny}
Figure~\ref{fig:rhopmax-Hp-tau002-tau003-Ny} shows temporal evolution of the maximum density of solids (upper panels) and their scale height (lower panels) for runs with $\tau_s = 0.02,~ Z = 0.013$ (left panels) and $\tau_s = 0.03,~ Z = 0.005$ (right panels). The horizontal line in the upper panels marks the Hill density, as defined in Equation~\eqref{eq:rhoH}.

\begin{figure*}[ht!]
	\centering
	\includegraphics[width=\textwidth]{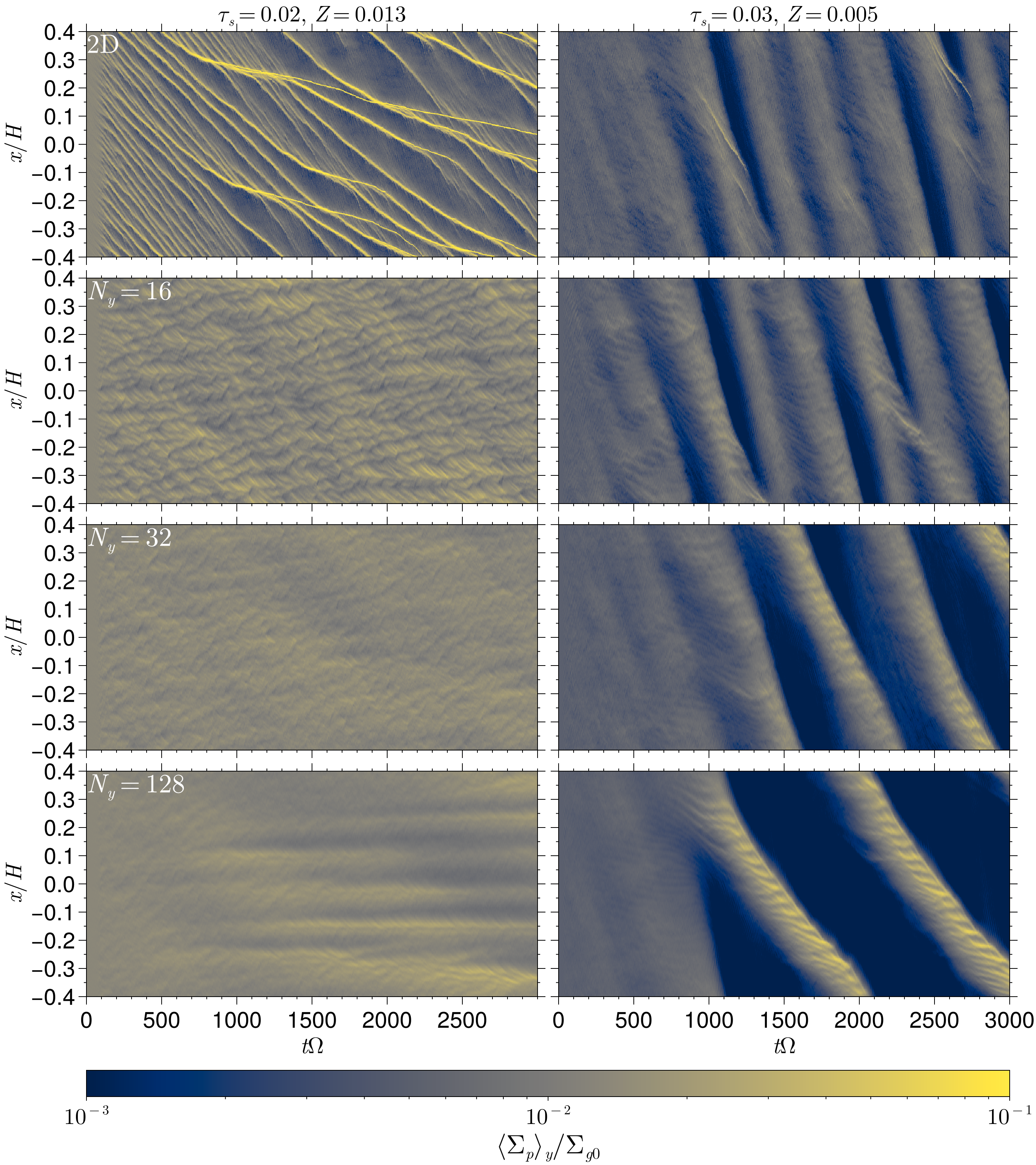}
	\caption{Evolution of dust filaments, shown as azimuthally averaged surface density $\langle \Sigma_p \rangle_y$ as a function of time and radial position $x$, for the simulations presented in Figure~\ref{fig:rhopmax-Hp-tau002-tau003-Ny}. The left and right columns correspond to $\tau_s = 0.02$, $Z = 0.013$ and $\tau_s = 0.03$, $Z = 0.005$, respectively. From top to bottom, panels correspond to simulations with $N_y = 1$ (2D), 16, 32, and 128 (fiducial 3D model). For $\tau_s = 0.02$, $Z = 0.013$, filaments in the 2D run drift inward and merge quickly, leading to strong clumping, whereas filaments in the 3D runs are much more quiescent. By contrast, for $\tau_s = 0.03$, $Z = 0.005$, dense filaments fail to form in all but the fiducial 3D run ($N_y = 128$), where a filaments-in-filaments structure is most pronounced.	
    }
    \label{fig:rhopt-tau002-tau003-Ny}
\end{figure*}

\begin{figure}
	\centering 
	\includegraphics[width=\columnwidth]{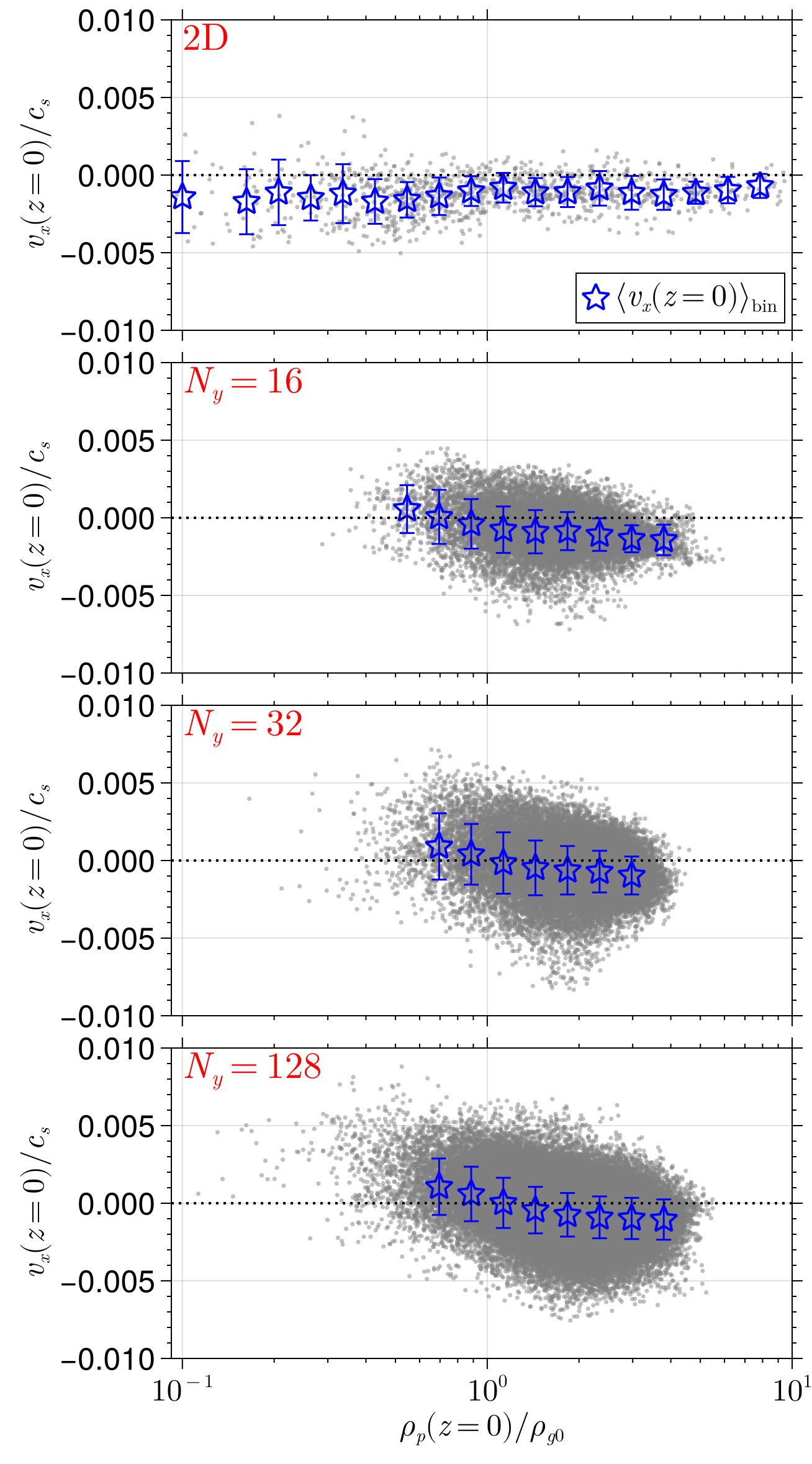}
	\caption{Scatter plots of radial velocity of solid particles $(v_x)$ versus their density $(\rho_p)$ at the midplane $(z=0)$ for $\tau_s=0.02,~Z=0.013$ with different values of $N_y$. The snapshots are taken at $t\Omega=1000$. We sample $v_x$ and $\rho_p$ from two horizontal slabs centered on the midplane, yielding 1024 and $\sim 1.3 \times 10^5$ data points for the 2D (left) and 3D ($N_y = 128$, right) simulations, respectively. Blue stars indicate the mean radial velocity within density bins, with error bars denoting the standard deviation. The bin average is plotted only when the bin contains more than 1\% of the total data points. In the 2D model, the radial drift speed slows down as $\rho_p$ increases, which can facilitate filaments merging. By contrast, the 3D models $(N_y>1)$ show a weak negative correlation between $v_x$ and $\rho_p$, with mean radial velocities becoming positive for $\rho_p(z=0)/\rho_{g0} \lesssim 1$.
	}
		\label{fig:rhop-vx-scatter-tau002-Z0013}
\end{figure}

\begin{figure*}[ht!]
	\centering
	\includegraphics[width=\textwidth]{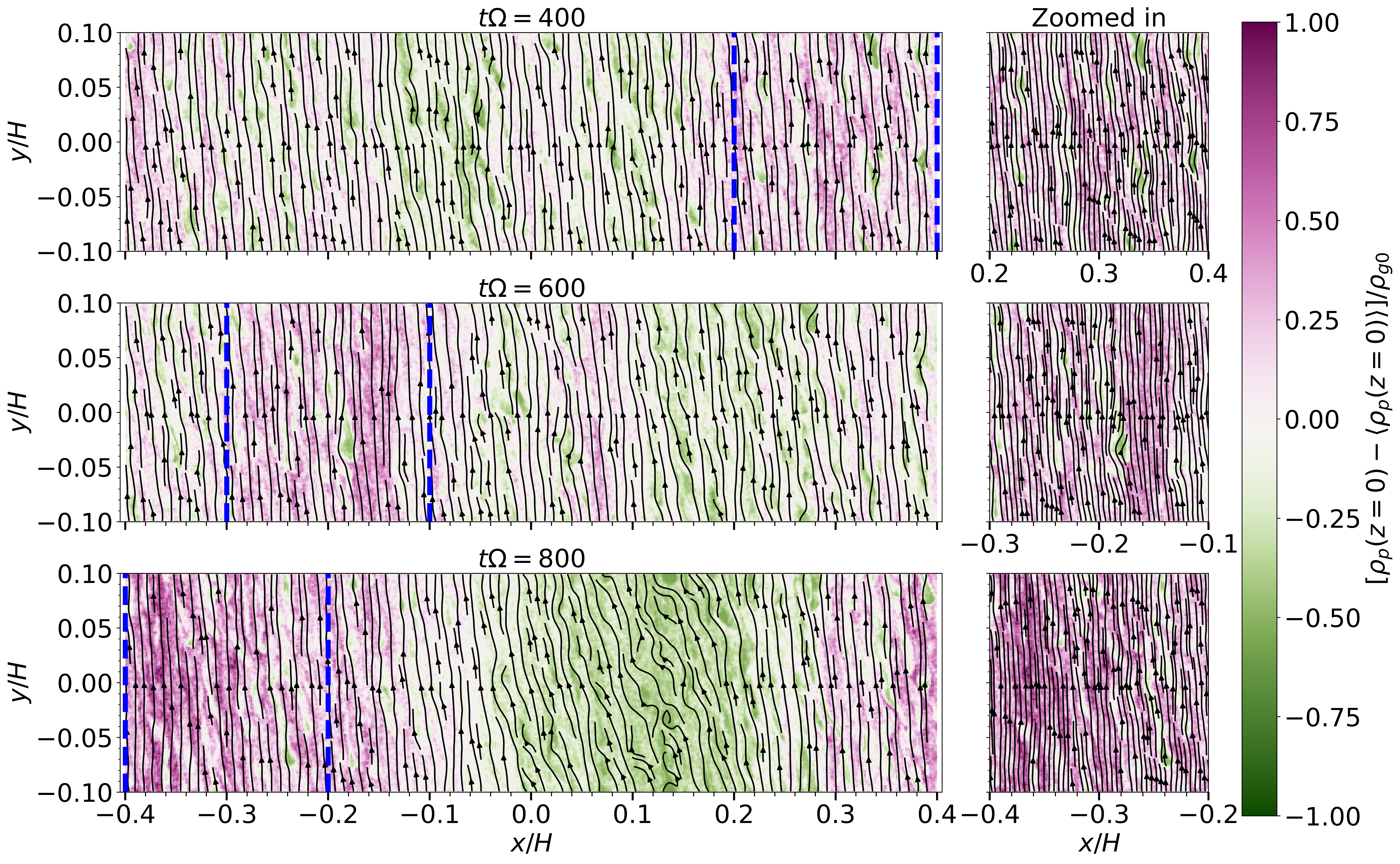}
	\caption{Streamlines of solid particles in the midplane, with background color indicating midplane density deviations from its mean value for the case with $\tau_s = 0.03$ and $Z = 0.005$ at $N_y=128$. From top to bottom, the snapshots correspond to $t\Omega = 400$, 600, and 800, respectively. The right-hand panels provide a zoomed-in view of the high-density region bounded by the two blue vertical lines in the left panels, illustrating particle motion within filaments at higher resolution. The motion of particles and their density distribution are clearly non-axisymmetric. This asymmetry induces radial motion of particles within the high-density region, allowing some to drift inward and redistribute material throughout the region. Simultaneously, the high-density region is fed by particles from the surrounding low-density areas. This process can give rise to the filaments-in-filaments structure observed in, for example, Figure~\ref{fig:rhopt-tau002-tau003-Ny} (see text for details).}
	\label{fig:tau003-Z0005-streamline}
\end{figure*}

The effect of $N_y$ on the maximum density is evident and acts in opposite directions for the two parameter sets. In the $\tau_s = 0.02$ runs, only the 2D model ($N_y = 1$) shows strong clumping. All 3D models ($N_y > 1$) exhibit similar maximum densities, which are more than an order of magnitude lower than that of the 2D case. By contrast, in the $\tau_s = 0.03$ runs, only the fiducial 3D model ($N_y = 128$) produces strong clumping. The other three runs have comparable maximum densities, although the 2D model occasionally reaches higher values.

In addition, the azimuthal dimension influences particle scale height, $H_p^*$. In the case of $\tau_s = 0.02$, the 2D model exhibits the largest $H_p^*$, whereas all three 3D models have similarly smaller values of $H_p^*$. A similar trend is observed in the $\tau_s = 0.03$ runs and is even more pronounced: $H_p^*$ decreases with increasing $N_y$. This trend of smaller scale heights in 3D simulations was also reported in previous studies \citepalias{Yang2017,LiYoudin21}. In general, a smaller $H_p^*$ leads to higher $\epsilon$ and thus more easily facilitates strong clumping. While this correlation holds for the $\tau_s = 0.03$ cases, the $\tau_s = 0.02$ cases deviate from this trend, showing weaker clumping despite having lower $H_p^*$ in 3D. We note that, however, the difference in $H_p^*$ between 2D and 3D models is smaller in the $\tau_s=0.02$ cases than in the $\tau_s=0.03$ cases.

\begin{figure}[ht!]
    \centering
    \includegraphics[width=\columnwidth]{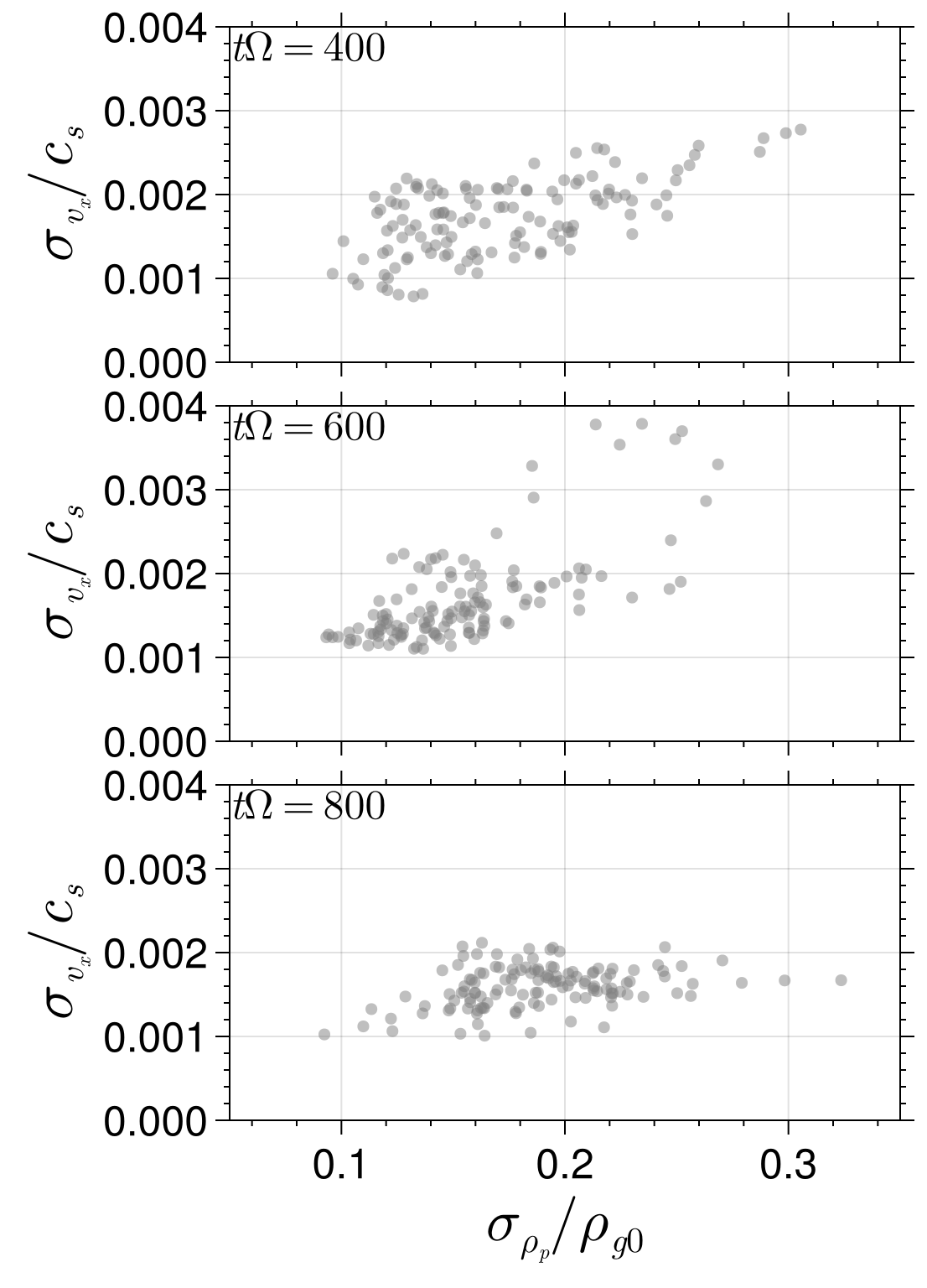}
    \caption{Scatter plots of the standard deviations of $v_x(x,y,z=0)$ ($\sigma_{v_x}$) and $\rho_p(x,y,z=0)$ ($\sigma_{\rho_p}$), computed by taking the standard deviation along the $y$-direction at each $x$. These quantities are used to quantify the variation of radial velocity and particle density along $y$. We show data points from the zoomed-in region in the right panels of Figure~\ref{fig:tau003-Z0005-streamline}, focusing on the correlation within the high-density region. A positive correlation is observed between $\sigma_{v_x}$ and $\sigma_{\rho_p}$, though it is weaker at $t\Omega = 800$. This suggests that radial velocity fluctuations along $y$ are more pronounced when solids are embedded in regions of non-axisymmetric density structure.
    } 
    \label{fig:tau003-Z0005-sigmarhop-sigmavx}
\end{figure}

To examine how varying $N_y$ affects formation of dust filaments, we present spacetime plots of $\langle \Sigma_p \rangle_y$\footnote{By examining space-time diagrams for specific $y$-slices, we have confirmed that the morphology of the filaments shown in Figure~\ref{fig:rhopt-tau002-tau003-Ny} does not vary with $y$; that is, the averaging procedure does not remove important information.} (i.e., the quantity being plotted versus time and $x$) for $\tau_s=0.02$ cases (left) and $\tau_s=0.03$ cases (right) in Figure \ref{fig:rhopt-tau002-tau003-Ny}. From top to bottom, panels correspond to runs $N_y=1$, 16, 32, and 128, respectively. The figure reveals a striking difference in the evolution of filaments across runs with different values of $N_y$ within each parameter set. Moreover, the effect of $N_y$ varies between the two parameter sets. We explain these results in detail below.

\subsubsection{$\tau_s = 0.02$}\label{sec:results:comp_2D:Ny:tau002}
In the $\tau_s = 0.02$ cases (left panels in Fig.~\ref{fig:rhopt-tau002-tau003-Ny}), filaments in the 2D model drift radially inward and undergo merging over time. These merging events become particularly pronounced after $t\Omega \sim 500$, driving $\rhopmax$ above the Hill density (see Fig.~\ref{fig:rhopmax-Hp-tau002-tau003-Ny}). In contrast to the 2D model, filaments in the three 3D models are less prominent and exhibit neither significant radial drift nor merging. We note, however, that individual particles drift inward due to gas drag, and no pressure trap is present in our simulations. In other words, the stationary filaments represent a pattern motion that is decoupled from the actual movement of the material—that is, the individual particles. Overall, temporal evolution of filaments in the 3D runs is much more quiescent, consistent with lower $\rhopmax$ compared to the 2D model.

To better understand the different behavior of filaments for different values of $N_y$, Figure \ref{fig:rhop-vx-scatter-tau002-Z0013} presents correlations between radial velocity $(v_x)$ and density $(\rho_p)$ of solid particles at the midplane at $t\Omega=1000$. From top to bottom, panels correspond to $N_y=1,~16,~32$, and 128, respectively. We mark mean radial velocities within density bins as blue stars. 

The radial motion of solid particles in the 2D model remains predominantly inward, as indicated by most bin-averaged velocities lying below the horizontal line at $v_x(z=0) = 0$. The mean radial velocity of solids decreases as $\rho_p(z=0)$ increases, meaning that solid particles inside dense filaments drift inward more slowly than those outside. This behavior allows solids to continually feed into filaments, promoting strong clumping.

In contrast to the 2D case, where solid particles drift radially inward regardless of whether they are inside or outside filaments, the 3D models (bottom three panels of Fig.~\ref{fig:rhop-vx-scatter-tau002-Z0013}) show 
radial velocity distributions that are broader at all values of $\rho_p(z=0)$. This suggests enhanced stirring in 3D compared to 2D, and as part of this turbulence, there is more outward radial motion of particles in 3D compared to 2D, especially when $\rho_p(z=0)/\rho_{g0} < 1$. This increased stirring thus reduces the net flux of solid particles into filaments, leading to the stationary motion of the filaments seen in the left panels of Figure~\ref{fig:rhopt-tau002-tau003-Ny} as individual particles continuously enter and exit the filaments\footnote{This behavior is analogous to a traffic jam: if cars enter and leave the jam at the same rate, the traffic jam remains stationary.}. Since filaments do not drift inward in the 3D cases, merging between them is suppressed. This inhibits strong clumping and leads to significantly lower values and a more quiescent evolution of $\rhopmax$ compared to the 2D case (Fig.~\ref{fig:rhopmax-Hp-tau002-tau003-Ny}).

\subsubsection{$\tau_s = 0.03$}\label{sec:results:comp_2D:Ny:tau003}

We now present results for the $\tau_s = 0.03$ cases. As shown in the right panels of Figure~\ref{fig:rhopt-tau002-tau003-Ny}, a filaments-in-filaments structure becomes increasingly pronounced as $N_y$ increases from 1 (2D) to 128 (fiducial 3D). In the 2D model, thin and dense filaments occasionally appear, briefly driving $\rhopmax$ above that in the $N_y = 16$ and 32 runs (see Fig.~\ref{fig:rhopmax-Hp-tau002-tau003-Ny}). However, these filaments do not persist. This comparison suggests that the filaments-in-filaments structure is an intrinsically non-axisymmetric phenomenon. In what follows, we provide a qualitative explanation for the onset of this structure.

Figure~\ref{fig:tau003-Z0005-streamline} show the streamlines of solid particles at the midplane for the fiducial 3D run ($N_y=128$). The background color indicates the deviation of their density from its spatial average: $\rho_p(z=0) - \langle \rho_p(z=0) \rangle$. At each snapshot, the right panel provides a zoomed-in view of the region bounded by the two blue vertical lines in the left panel. 

As shown in the figure, the fiducial 3D run clearly exhibits non-axisymmetric motion. In other words, solid particles show streamlines bent radially inward or outward along the azimuthal direction (see the right panels; see also \citetalias{Yang2017} for a similar phenomenon). Furthermore, the radial velocity varies more significantly along $y$ in filaments that exhibit strong density variations in the same direction. 

For more quantitative evidence for the claimed correlation, we analyze the midplane radial velocity $v_x(x,y)$ and particle density $\rho_p(x,y)$ shown in Figure~\ref{fig:tau003-Z0005-streamline} by computing the standard deviation of each quantity at a given $x$. These standard deviations quantify the extent to which radial velocity and particle density vary along the $y$-direction at each radial location. Figure~\ref{fig:tau003-Z0005-sigmarhop-sigmavx} shows the standard deviations of the midplane radial velocity ($\sigma_{v_x}$) and particle density ($\sigma_{\rho_p}$). To focus on high-density regions, we show these values within the zoomed-in area highlighted in the right panels of Figure~\ref{fig:tau003-Z0005-streamline}. Larger values of $\sigma_{\rho_p}$ are associated with increased $\sigma_{v_x}$, suggesting that stronger density variations in $y$ lead to more pronounced radial velocity variation of particles.

Overall, Figures~\ref{fig:tau003-Z0005-streamline}–\ref{fig:tau003-Z0005-sigmarhop-sigmavx} suggest that within high-density regions (e.g., outlined by the blue dashed lines in Fig.~\ref{fig:tau003-Z0005-streamline}), particles can still drift radially inward due to the density gradient along the $y$-direction. This inward flow within high-density regions is caused by non-axisymmetry and appears to contribute to the distinguishing structure of filaments between the 2D and 3D cases; thin filaments appear and dissolve in 2D, whereas in 3D, multiple closely spaced filaments (i.e., the filaments-in-filaments structure) form and persist. In what follows, we describe the onset of this structure based on our analysis of the data. 

In both 2D and 3D, filaments emerge from high-density regions (see $t\Omega \approx 1000$ in the top and bottom right panels of Fig.~\ref{fig:rhopt-tau002-tau003-Ny} for the 2D and 3D cases, respectively). However, in 3D, non-axisymmetric radial motion allows some particles to drift inward even within dense regions, although this drift is slowed by the high values of $\rho_p$. Simultaneously, upstream particles (those in the low-density region shown by green shading in Fig.~\ref{fig:tau003-Z0005-streamline}) catch up and enter the high-density region. Since these upstream particles drift faster than those within the high-density region, the local density there can grow over time. In addition, the inward motion of particles within the high-density region helps redistribute material across its radial extent. Together, this inward flow and the accumulation of upstream particles enable the formation of multiple filaments. Once the density becomes sufficiently high, multiple narrowly spaced filaments form within the high-density region (in the 3D case, this occurs at $t\Omega\approx 1000$; see bottom right in Fig.~\ref{fig:rhopt-tau002-tau003-Ny}). These filaments, embedded within the broader high-density region (i.e., the parent filament),  drift slowly inward and occasionally merge over time.

In 2D, the mechanism described above is suppressed. Although the radial motion of solids slows down within high-density regions (as in 3D), the absence of azimuthal density variations (i.e., $\partial \rho_p / \partial y = 0$) confines particles to predominantly azimuthal motion. As a result, particles accumulate within a narrow radial extent rather than spreading across a broader region. This leads to the formation of thin filaments in 2D. Over time, such filaments could grow in density by collecting particles from upstream. Although filaments fail to grow in our 2D case, this behavior is broadly consistent with the axisymmetric simulations of \citetalias{LiYoudin21}, where overdensities appear as thin filaments rather than the filaments-in-filaments structure seen in our 3D simulations. In short, the ability of particles to escape individual filaments in the lower density regions of those filaments allows for the more complex structure that we see in our simulations.

\begin{figure*}[ht!]
	\centering
	\includegraphics[width=\textwidth]{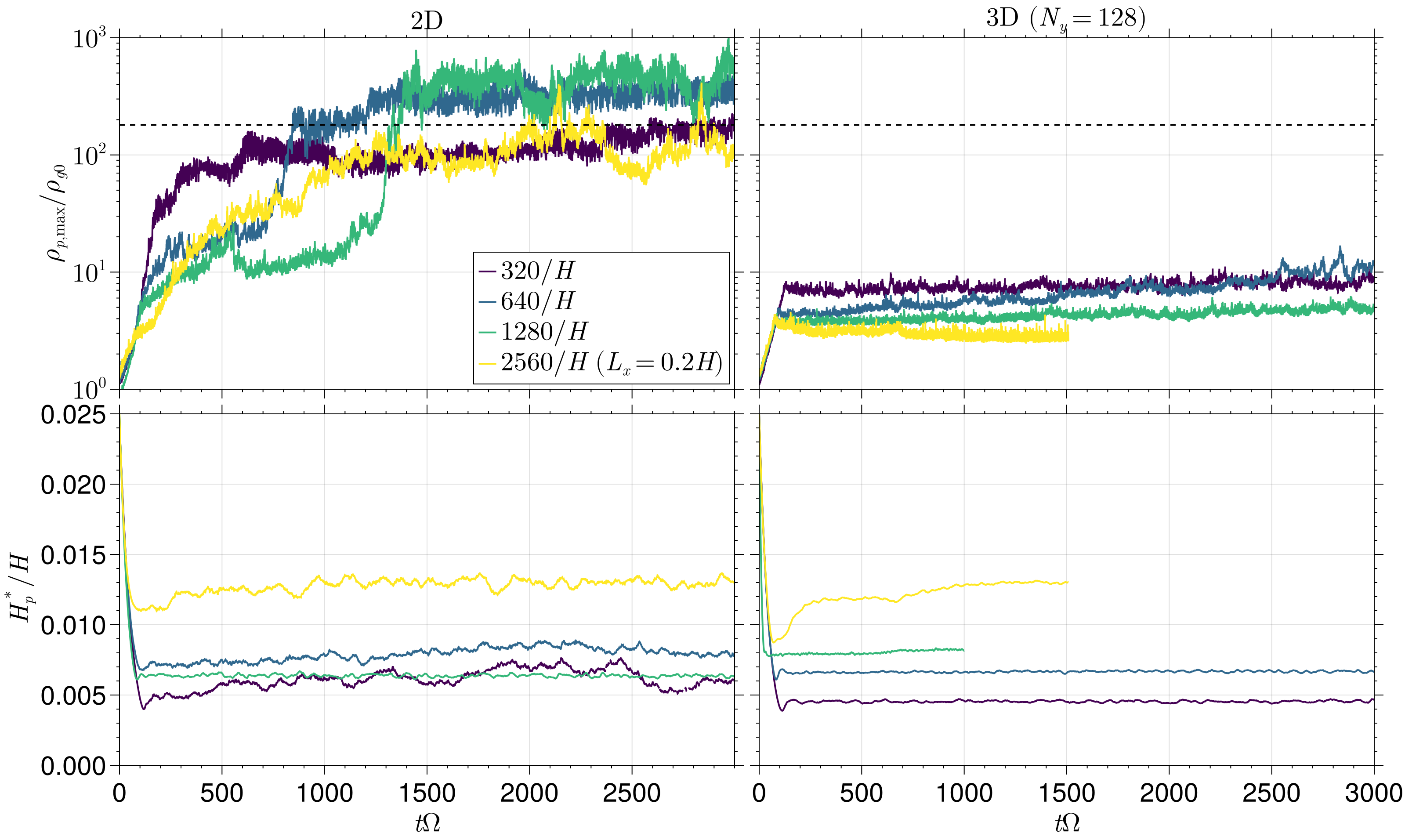}
	\caption{Similar to Figure \ref{fig:rhopmax-Hp-tau002-tau003-Ny}, but for the resolution test at $\tau_s=0.02,~Z=0.013$. Left and right panels correspond to 2D and 3D models, respectively. In each plot, darker to lighter colors denote grid resolutions of $320/H$ to $2560/H$. The smaller radial domain of $0.2H$ is used for the highest resolution. At any given resolution, strong clumping occurs in 2D models, whereas in 3D models, $\rhopmax$ decreases with increasing resolution. The effect of resolution on $H_p^*$ is the same in 2D and 3D models: it increases with increasing resolution. 
	}
    \label{fig:rhopmax-Hp-tau002-res}
\end{figure*}
\subsubsection{Interpretation of our results}\label{sec:results:comp_2D:Ny:interpretation}
In this section, we summarize our results for the two $\tau_s$ cases (Sections~\ref{sec:results:comp_2D:Ny:tau002}-\ref{sec:results:comp_2D:Ny:tau003}). We then explain how these results lead to different $\Zcrit$ in 3D compared to the previous 2D simulations as well as the sharp transition in $\Zcrit$ observed between $\tau_s = 0.02$ and 0.03 (Fig.~\ref{fig:Zcrit}).

For $\tau_s = 0.02$ and $Z = 0.013$, strong clumping occurs only in 2D (Fig.~\ref{fig:rhopmax-Hp-tau002-tau003-Ny}, top left). In 2D, filaments drift inward and merge, while in 3D they appear stationary and fail to grow (Fig.~\ref{fig:rhopt-tau002-tau003-Ny}). This difference arises from enhanced stirring in 3D (Fig.~\ref{fig:rhop-vx-scatter-tau002-Z0013}), which drives significant outward motion in low-density regions and reduces the net influx of solids into filaments, thereby suppressing strong clumping.

In contrast, for $\tau_s = 0.03$ and $Z = 0.005$, strong clumping occurs only in the 3D model with $N_y = 128$. This is due to a filaments-in-filaments structure that becomes more pronounced at larger $N_y$ (Fig.~\ref{fig:rhopt-tau002-tau003-Ny}, right panels). This complex structure is driven by the azimuthal variations in $\rho_p$ that induces radial flows within a high-density region (Figs.~\ref{fig:tau003-Z0005-streamline}-\ref{fig:tau003-Z0005-sigmarhop-sigmavx}).

These results show that the filament behavior in 3D transitions from a filaments-in-filaments structure to a more quiescent, stationary configuration with suppressed radial motion as $\tau_s$ decreases and $\epsilon$ increases from below to above 1.\footnote{The linear theory of the SI predicts distinct behavior across the $\epsilon < 1$ and $\epsilon > 1$ regimes. For instance, the peak growth rate increases sharply as $\epsilon$ exceeds 1 for $\tau_s \ll 1$ \citep{YJ07,JY07}. Additionally, the SI transitions from resonant drag behavior when $\epsilon < 1$ to non-resonant behavior when $\epsilon > 1$ \citep{SquireHopkins18}. However, we do not directly connect these linear results to our findings, as the linear theory is shown to be a poor predictor of strong clumping \citepalias{LiYoudin21,Lim25}.} In fact, as $\tau_s$ decreases and $\epsilon$ increases, solid particles become more tightly coupled to the gas and exert stronger backreaction, leading to slower radial drift. Although this is true in both 2D and 3D setups, the azimuthal dimension seems to amplify the effect in 3D simulations by enhancing stirring. Consequently, while filaments consistently drift inward in 2D simulations regardless of $\epsilon$ and $\tau_s$ (e.g., \citetalias{LiYoudin21}), their radial motion in 3D becomes increasingly suppressed as $\tau_s$ decreases and $\epsilon$ increases. As a result, compared to 2D, $\Zcrit$ in 3D is lower for $\tau_s \geq 0.03$ and $\epsilon < 1$, where a filaments-in-filaments structure promotes strong clumping, but higher for $\tau_s \leq 0.02$ and $\epsilon > 1$, where the lack of radial filament motion inhibits strong clumping. 

Finally, since the sharp transition in $\Zcrit$ coincides with $\epsilon$ increasing from below to above 1 (Figs.~\ref{fig:Zcrit}–\ref{fig:epscrit}), the complex filament behavior—varying with $\tau_s$ and $\epsilon$—is likely responsible for this transition. Specifically, a filaments-in-filaments structure promotes strong clumping for $\tau_s \geq 0.03$, where $\epsilon < 1$ (Fig.~\ref{fig:rhopt-Z0004}). However, as $\tau_s$ decreases to 0.02 and $\epsilon$ exceeds 1, the radial motion of filaments becomes suppressed. Consequently, much higher values of $Z$ are required to achieve strong clumping in this regime, where filament merging becomes infrequent or absent. By contrast, the relatively simple and consistent behavior of filaments in 2D may explain the smooth, monotonic trend in $\Zcrit$ reported by \citetalias{Lim25}. Notably, they resolved the sharp transition in $\Zcrit$ originally reported by \citetalias{LiYoudin21} by increasing grid resolution. Therefore, in the next subsection, we investigate whether resolution affects the sharp transition observed in our $\Zcrit$.

\begin{figure*}[ht!]
	\centering
	\includegraphics[width=\textwidth]{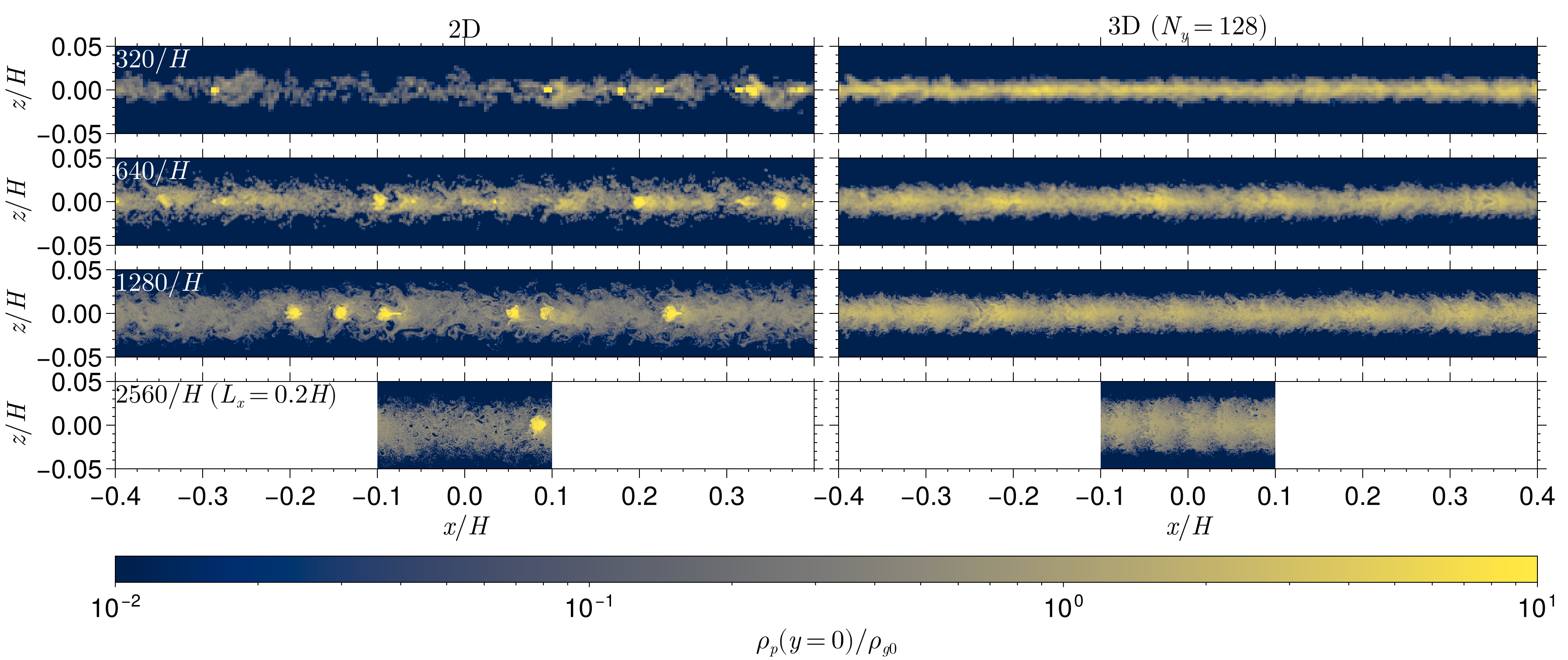}
	\caption{Final snapshots of density of solids at $y=0$ for 2D (left) and 3D (right) models. From top to bottom, panels correspond to simulations with resolutions of $320/H$, $640/H$, $1280/H$, and $2560/H$. The vertical domain is zoomed in to $\pm 0.05H$ around the midplane, though the full vertical extent is $0.2H$. In both 2D and 3D cases, the dust layer becomes thicker with increasing resolution. However, dense clumps at the midplane appear only in the 2D models.	
		}
		\label{fig:snapshots-tau002-res}
\end{figure*}

\subsection{Effect of Grid Resolution}\label{sec:results:comp_2D:resolution}
The degree of concentration driven by the SI is known to be sensitive to grid resolution as previous studies found that higher resolution yields stronger clumping (\citealt{YangJohansen14}; \citetalias{Yang2017,Lim25}). For example, \citetalias{LiYoudin21} reported a sharp increase of their $\Zcrit$ between $\tau_s=0.01$ and 0.02, which \citetalias{Lim25} later demonstrated was a resolution effect (see Section~\ref{sec:results:crit_ratio:Zcrit} for details).

Motivated by this, we perform a resolution test at $(\tau_s, Z) = (0.02, 0.013)$ for both 2D and 3D simulations—a particularly relevant case since this $Z$ value lies just below our clumping threshold at $\tau_s = 0.02$, and the parameter set resides on the low-$\tau_s$ side of the sharp rise in $\Zcrit$. We consider resolutions up to $2560/H$. While we maintain the same domain size across resolutions up to $1280/H$, we use a smaller radial domain ($L_x = 0.2H$) for the highest-resolution runs to reduce computational cost. As we show in the remainder of this subsection, SI-induced particle concentration strengthens with increasing resolution in 2D, whereas it weakens in 3D. Although the physical mechanism underlying this contrasting resolution dependence remains unclear in this work, we highlight several important trends that may help elucidate this behavior.

Figure~\ref{fig:rhopmax-Hp-tau002-res} presents the time evolution of the maximum density of solids and their scale heights in the upper and lower panels, respectively. The results of 2D and 3D models are shown in the left and right panels, respectively. In each panel, darker to lighter colors indicate increasing resolution. 

First, the effect of resolution differs significantly between 2D and 3D models. In 2D, the saturated maximum density increases with resolution from $320/H$ to $1280/H$. The lowest resolution run ($320/H$) barely reaches the Hill density (marked by the horizontal line), whereas the $640/H$ and $1280/H$ runs exceed it by a factor of a few. The highest resolution run ($2560/H$) shows somewhat lower maximum densities than the intermediate-resolution runs. This is likely due to its smaller radial domain, which reduces the number of filaments formed (see Fig.~\ref{fig:snapshots-tau002-res}) and thus decreases the frequency of filament merging. Nevertheless, all 2D simulations across resolutions exhibit strong clumping. In contrast, none of the 3D runs show strong clumping, with maximum density remaining below $\approx 10\rho_{g}$. Furthermore, the maximum density slightly decreases with increasing resolution.  

Second, in contrast to the resolution dependence of $\rhopmax$, the scale height of solids increases with resolution in both 2D and 3D models as the bottom panels of Figure~\ref{fig:rhopmax-Hp-tau002-res} show. Notably, the largest increase in scale height occurs between the $1280/H$ and $2560/H$ runs in both cases. This trend suggests that resolving smaller-scale turbulent motions at higher resolution enhances vertical stirring, leading to a thicker dust layer, which is also reported in previous 2D simulations \citepalias{LiYoudin21,Lim25}. 

In Figure~\ref{fig:snapshots-tau002-res}, we show final snapshots of $\rho_p$ at $y = 0$ for the simulations presented in Figure~\ref{fig:rhopmax-Hp-tau002-res}. In the 2D simulations, dense clumps are clearly visible at the midplane. The highest-resolution run forms only a single clump, while the lower-resolution runs exhibit multiple clumps. Frequent filament merging in these lower-resolution cases (see the top left panel of Fig.~\ref{fig:rhopt-tau002-tau003-Ny} for the $640/H$ case) contributes to their higher maximum densities compared to the highest-resolution run. By contrast, no clumps form in any of the 3D simulations.

Finally, the morphology of dust layer varies with resolution and between 2D and 3D. In both 2D and 3D cases, the higher resolution is the thicker the dust layer becomes, consistent with the trend of $H_p^*$ with resolution seen in Figure \ref{fig:rhopmax-Hp-tau002-res}. In addition, low-density cavities are visible in both 2D and 3D at the highest resolution. Furthermore, at any given resolution, the top of the dust layer is more corrugated in 2D than in 3D (see also \citetalias{Yang2017}). Interestingly, in the 3D cases, the dust layer in the highest-resolution run exhibits a well-separated, band-like structure across the radial domain, in contrast to the smoother and more uniform layers seen in the lower-resolution runs. However, since $L_x$ is different in the highest-resolution run, it remains unclear if the morphological difference is due to increase in resolution or decrease in radial domain sizes. 

In conclusion, the sharp jump in our $\Zcrit$ curve persists even when the resolution is increased by a factor of four from the fiducial value ($640/H$). As discussed in \citetalias{Lim25}, increasing resolution has two competing effects: it enhances clumping but also triggers greater stirring. Based on the results presented here, the former dominates in the 2D simulations with $(\tau_s,Z)=(0.02,0.013)$, while the latter dominates in the corresponding 3D cases. Furthermore, \citetalias{Lim25} identified a critical resolution of $2560/H$ for the transition from no clumping to strong clumping in 2D simulations with $\tau_s = 0.01$. Given the comparable value of $\tau_s$ and the fact that the 3D run with $2560/H$ attains a low degree of clumping ($\rho_{p,\textrm{max}}/\rho_{g0} \approx 1$), the case of $\tau_s = 0.02$, $Z = 0.013$ may not transition to strong clumping even at resolutions beyond $2560/H$.

\section{Discussion}\label{sec:discussion}
\subsection{Implications for Planetesimal Formation via the SI}\label{sec:discussion:planetesimals}
The results presented in Sections~\ref{sec:results_3D}–\ref{sec:results:comp_2D} show that for $\tau_s \gtrsim 0.03$, strong clumping can occur at significantly lower critical $Z$ values—down to 0.002 at $\tau_s~=~0.1$—compared to those reported in previous 2D simulations \citepalias{carrera_how_2015,Yang2017,LiYoudin21,Lim25}. By contrast, for $\tau_s~=~10^{-3}$–$10^{-2}$, our 3D simulations require higher $Z$ values than those found in earlier 2D studies. This implies that planetesimal formation via the SI either requires substantial grain growth to reach the optimal $\tau_s$, or in the absence of such growth, a solids-to-gas column density ratio exceeding the canonical ISM value of 0.01 is needed.

Recent observations of circumstellar disks suggest that planetesimal and planet formation may already be underway during the Class 0/I phases. For instance, several studies of Class I sources report evidence of rapid grain growth to centimeter sizes—corresponding to $\tau_s \approx 0.1$ at 5 au \citep{youdin_grains_2010}—indicating significant dust evolution during these early stages \citep{Zhang21,Han23,Han25,Radley25}. In parallel, high-resolution ALMA images have revealed the ubiquity of dust rings in disks (see e.g.,\citealt{ALMA2015,andrews_disk_2018}) including in some of the youngest Class 0/I systems \citep{segura-cox_four_2020}. If planets are responsible for these ring structures, then our finding that the SI can trigger strong clumping at low $Z$ values for cm-sized solids suggests that the SI may play a key role in forming the first generation of planetesimals during the earliest stages of disk evolution.

In addition, we find that $H_p^*$ decreases within dense filaments formed by the SI (Figure \ref{fig:rhop-Hpeff}), indicating that the velocity dispersion of solids is reduced in these regions. This can facilitate further dust growth within the filaments. Indeed, recent numerical studies have shown that the interplay between dust coagulation and the SI enables grains to grow in size, thereby reaching the conditions for planetesimal formation more easily than the SI alone \citep{Tominaga_Tanaka23,Ho+24,Carrera25_positive_feedback,Tominaga_Tanaka25}. This SI-assisted grain growth presents a promising pathway to planetesimal formation by growing solids in SI-driven filaments toward the optimal regime. 


Despite this, it remains debated whether the SI can drive planetesimal formation in disks with more realistic levels of turbulence—those exceeding the turbulence generated by the SI itself. On the one hand, \citet{Morbidellis22} modeled a disk undergoing infall from a collapsing molecular cloud and found that planetesimal formation via the SI may occur within the first 0.5 Myr due to significant dust pile-up at both the water snowline and the silicate sublimation line. Similarly, \citet{Kawasaki25} studied dust growth in Class 0/I disks including both infall and MHD winds. They found that planetesimal formation via the SI is more likely to occur with MHD winds than without (see their Fig.~17). However, we caution that both studies assumed a globally turbulent disk but used SI clumping thresholds that do not account for the effects of turbulence. Specifically, \citet{Morbidellis22} adopted a threshold of $\epsilon > 0.5$ for planetesimal formation, whereas \citet{Lim24} demonstrated that significantly higher $\epsilon$ values are needed when external turbulence is present (see their Fig.~4). Likewise, \citet{Kawasaki25} used the clumping threshold from \citetalias{LiYoudin21}, which was derived from simulations in which no turbulence was present other than that induced by the SI.

On the other hand, dust evolution models that considered the effect of turbulence on the clumping threshold \citep{Estrada23,Carrera25_dust_growth}\footnote{While \citet{Estrada23} adopted the clumping threshold from \citetalias{LiYoudin21}, they modified the criterion to incorporate the effect of external turbulence (see Equations 13–14 in \citetalias{LiYoudin21}) and applied it to their analysis.} have shown that during the first Myr of disk evolution, dust growth and concentration do not meet any of the SI criteria established in previous studies (\citetalias{carrera_how_2015,Yang2017,LiYoudin21}; \citealt{Lim24}; \citetalias{Lim25}). Moreover, turbulence is expected to suppress the growth of the SI and hinder planetesimal formation, as demonstrated by both linear stability analyses \citep{Chen20,UC20} and numerical simulations with externally driven turbulence \citep{Gole20,Lim24}. 

However, turbulence driven by (magneto)hydrodynamical instabilities—such as the magnetorotational instability (MRI; \citealt{BH92,Hawley95}) and the vertical shear instability (VSI, \citealt{Nelson13VSI})—is shown to enhance solid concentration either by creating weak pressure bumps \citep{Schaffer2021,HB25b} that facilitate the onset of the SI or by forming dust traps \citep{Johansen06,xu_dust_2022} that halt radial drift entirely. 

This all being said, most of these studies only explored $\tau_s=0.1$ or even larger. Moreover, although limited to the outer regions of PPDs, observational evidence supports the presence of weak turbulence \citep{Flaherty15, Pinte16, DUllemond18, Flaherty18, Flaherty20, Villenave22}. In addition, turbulence can be further suppressed near the midplane if solids exert sufficient mass-loading the gas \citep{xu_turbulent_2022,Lim24,HB25a}, which could enhance further settling and trigger the SI. 

Overall, despite recent observations and theoretical progress incorporating additional physics (e.g., dust coagulation or turbulence), the ability of the SI to form planetesimals remains an area of active study. One of the outstanding issues is whether and how various instabilities, such as the MRI and VSI, operate under realistic conditions in circumstellar disks. Simulations of global 3D disks that include comprehensive physics—such as radiation, gas self-gravity, and non-ideal MHD effects—can provide more realistic levels of turbulence throughout the disk. These turbulence profiles can then be incorporated into 1D disk evolution models to simulate dust growth under more realistic disk conditions. Our results can then serve as a benchmark to estimate the local solid-to-gas ratio ($Z$ or $\epsilon$) required to trigger planetesimal formation across a wide range of $\tau_s$ values if turbulence is weak near the disk midplane.

\subsection{Caveats}\label{sec:discussion:caveats}
There are a number of uncertainties and caveats associated with our results. First of all, we neglect the self-gravity of solids and instead use the Hill density (Equation~\ref{eq:rhoH}) as a proxy to assess whether clumping is sufficient for gravitational collapse. However, the Hill density is a necessary but not sufficient condition. For example, it is possible for turbulent diffusion to counteract self-gravity even if the Hill density is reached \citep{Gerbig20,klahr_turbulence_2020,gerbig_planetesimal_2023}. Thus, including self-gravity is necessary to confirm whether strong clumping indeed leads to planetesimal formation, particularly near the clumping threshold. 

In addition, while our simulations are fully 3D, the fiducial grid resolution is lower than that used in prior 2D studies \citepalias{Yang2017,LiYoudin21,Lim25}. However, our resolution test in Section~\ref{sec:results:comp_2D:resolution}, which targets the region where $\Zcrit$ rises sharply, does not show a transition from no clumping to strong clumping. This suggests that grid resolution may not significantly affect the clumping threshold, though future high-resolution studies are needed to test the convergence of the threshold. 

Another uncertainty lies with the effects of the domain size—particularly $L_x$ and $L_y$, which have been shown to influence filament formation (\citealt{YangJohansen14}; \citetalias{Yang2017}; \citealt{Li18,Schafer2024}; \citetalias{Lim25}). Our choice of $L_x$ is sufficiently large to capture filament interactions in that direction, as previous studies suggested that $L_x \gtrsim 0.4H$ is required for $\Pi = 0.05$ (\citealt{YangJohansen14,Li18}; \citetalias{Lim25}). However, since \citet{Schafer2024} demonstrated that the azimuthal extent of filaments is $\approx H$, our azimuthal domain size of $L_y = 0.2H$ is likely too small to accommodate more than a single filament along the $y$-direction. Although \citet{YangJohansen14} showed that strong clumping persists up to $L_x = L_y = 1.6H$ for $\tau_s = \pi/10$ and $Z = 0.02$, future studies should adopt sufficiently large $L_x$ and $L_y$ to fully capture filament interactions in both directions and  test the robustness of the $\Zcrit$ values found in this work.

Finally, our simulations assume a single solid size, whereas solids in disks are expected to exhibit a size distribution. The role of size distributions has been explored in both the linear \citep{Krapp2019,ZhuYang21} and nonlinear \citep{BaiStone10b_stratified,Schaffer2018,Schaffer2021,YangZhu2021,Rucska_Wadsley2023,Matthijsse2025} regimes of the SI. In particular, \citet{BaiStone10b_stratified} investigated strong clumping conditions using both 2D and 3D simulations with size distributions. They found that larger solids predominantly participate in clumping, consistent with the subsequent studies, and that the strong-clumping condition is more stringent in 3D than in 2D (see their Section 4 and Fig.~5). Specifically, among their four tested size distributions with $Z=0.03$—$\tau_s = 10^{-4}$–$10^{-1}$, $10^{-2}$–$10^{-1}$, $10^{-3}-10^0$, and $10^{-1}-10^0$—strong clumping in 3D occurred only for the $\tau_s = 10^{-1}$–1 case, whereas it occurred in all but the $\tau_s = 10^{-4}$–$10^{-1}$ case in 2D. This suggests that strong clumping in 3D favors size distributions dominated by larger solids compared to 2D. While this trend is broadly consistent with our findings that $\Zcrit$ in 3D is lower than in 2D for larger $\tau_s$, further work is needed to assess how significantly size distributions modify the clumping thresholds derived from single-size models, particularly by exploring a wider range of size distributions and $Z$ values.

\section{Summary}\label{sec:summary}
We explore conditions for strong clumping of solids by the SI with a suite of 3D local simulations of vertically stratified disks. In all of our simulations, the pressure gradient parameter $\Pi$ (Equation~\ref{eq:Pi}) is set to 0.05. Since we neglect the self-gravity of solids in our simulations, we use the Hill density (Equation~\ref{eq:rhoH}) to classify simulations into strong-clumping and no-clumping cases. A run is considered strong clumping if maximum density of solids in the simulation exceeds the Hill density. We provide the critical solid-to-gas column density ratio $(\Zcrit)$ and the density ratio at the midplane ($\epscrit$) as functions of their dimensionless stopping time $(\tau_s)$ above which strong clumping occurs in our simulations. In addition, we study the nature of SI-driven concentration and how it behaves differently from 2D axisymmetric (radial-vertical) simulations. Our main findings are as follows:

\begin{enumerate}
	\item We find that $\Zcrit$ reaches a minimum value of $\approx~0.002$ at $\tau_s=0.1$. As $\tau_s$ decreases from 0.1, $\Zcrit$ increases and exceeds 0.03 at $\tau_s = 10^{-3}$. However, we are unable to determine the precise value of $\Zcrit$ at $\tau_s = 10^{-3}$ because we have only a single simulation at this $\tau_s$. The clumping boundary is shown in Figure~\ref{fig:Zcrit}. 
	\item A sharp increase in $\Zcrit$ is found as $\tau_s$ decreases from 0.03 to 0.02, increasing from $\approx 0.005$ to $\approx 0.02$. Due to this sudden transition and the limited number of data points for $\tau_s \leq 0.02$, we only provide a fitting function to $\Zcrit$ for $\tau_s > 0.02$ (see Equation~\ref{eq:Zcrit}). We also conduct high-resolution simulations at $\tau_s = 0.02,~Z=0.013$ to assess whether resolution effects contribute to this transition. Even with a fourfold increase in resolution, the results remain unchanged (Section~\ref{sec:results:comp_2D:resolution}).
	\item Our 3D simulations yield values of $\Zcrit$ that differ significantly from those obtained in previous 2D simulations. For $\tau_s > 0.02$, we find lower $\Zcrit$ than the 2D-based criteria, while for $\tau_s \leq 0.02$, our $\Zcrit$ values are higher (Fig.~\ref{fig:Zcrit}). Correspondingly, we find $\epscrit < 1$ for $\tau_s > 0.02$ and $\epscrit > 1$ for $\tau_s \leq 0.02$ (Fig.~\ref{fig:epscrit}).
	\item The concentration of solids by the SI exhibits remarkably different behaviors between $\epsilon < 1$ and $\epsilon > 1$ (Section \ref{sec:results_3D:SI_clumping}). For $\epsilon < 1$, our 3D simulations show the emergence of subfilaments within larger parent filaments. This ``filaments-in-filaments” structure (see e.g., Fig.~\ref{fig:rhopt-Z0004})  is not observed in 2D simulations and appears to enhance clumping efficiency in 3D. We provide a qualitative explanation for the onset of the structure based on our analysis of the data in Section~\ref{sec:results:comp_2D:Ny}. 
	\item For $\epsilon > 1$, multiple filaments form across the radial domain and are evenly spaced, without developing the filaments-in-filaments structure. Although merging of filaments frequently occurs at $\tau_s = 0.1$, the radial drift speed of filaments decreases with decreasing $\tau_s$. This slower drift suppresses strong clumping when $\tau_s \leq 0.03$, as seen in our simulations. 
	\item However, 2D simulations show that filaments continue to drift radially inward even when $\epsilon > 1$, promoting mergers and enabling strong clumping. We show in Section \ref{sec:results:comp_2D:Ny:tau002} that 3D enhances stirring, causing a significant number of particles to move radially outward (Fig.~\ref{fig:rhop-vx-scatter-tau002-Z0013}). This in turn reduces the influx of solid particles into filaments, preventing filaments from growing in density and thus inhibiting strong clumping.
		\end{enumerate} 
In conclusion, our results demonstrate that the behavior of the SI is substantially different in 3D simulations compared to 2D axisymmetric simulations, leading to significantly different thresholds for strong clumping. These findings suggest that 2D simulations should be interpreted with caution, as they may not capture key physical processes inherent to fully three-dimensional systems.

\section*{Acknowledgments} 
We thank Chao-Chin Yang, Linn Eriksson, Daniel Carrera, and Andrew Youdin for fruitful discussions. J.L. acknowledges support from NASA under the Future Investigators in NASA Earth and Space Science and Technology grant \# 80NSSC22K1322. J.L. and J.B.S acknowledge support from NASA under Emerging Worlds grant \# 80NSSC20K0702. J.B.S. also acknowledges support from the National Science Foundation (NSF) under Grant No. AST-2407762. W.L. acknowledges support from the NASA Emerging Worlds program via grant \#80NSSC22K1419, and NSF via grants AST-2007422 and AST-2511672. J.L., J.B.S, and W.L. acknowledge support from the Theoretical and Computational Astrophysical Networks (TCAN) grant \# 80NSSC21K0497. The computations were performed using Stampede3 at the Texas Advanced Computing Center using XSEDE/ACCESS grant TG-AST120062 and on Pleiades at the NASA High-End Computing (HEC) Program through the NASA Advanced Supercomputing (NAS) Division at Ames Research Center. J.L. acknowledges the use of ChatGPT, which has been used on rare occasions to improve the clarity of the writing. The outputs have been checked for their correctness and no original content has been created.

\software{Julia \citep{bezanson2017julia}, Makie.jl \citep{DanischKrumbiegel2021}, Athena \citep{Stone08,stone_implementation_2010,BaiStone10a}, ChatGPT \citep{2023arXiv230308774O}}
\bibliography{main}{}
\bibliographystyle{aasjournalv7}

\end{CJK}
\end{document}